\documentclass[12pt]{article}
\usepackage{graphicx}
\usepackage{amsmath}
\usepackage{bm}
\def\Vec#1{\mbox{\boldmath $#1$}}
\def\mC{\mathcal C}

\def\itmb{\begin{itemize}}
\def\itme{\end{itemize}}
\def\enmb{\begin{enumerate}}
\def\enme{\end{enumerate}}
\def\eqnb{\begin{equation}}
\def\eqne{\end{equation}}

\def\PRL{Phys. Rev. Lett.}
\def\PRD{{Phys. Rev.} D}
\def\PRC{{Phys. Rev.} C}

\title{Triality selection rules of Octonion and \\Quantum Mechanics}  

\author{Sadataka Furui  \\
 Graduate School of Teikyo University\\
2-17-12 Toyosatodai, Utsunomiya, 320-0003 Japan {\thanks
{\textit{E-mail address:} furui@umb.teikyo-u.ac.jp}}
}

\begin{document}
\maketitle
\begin{abstract}%
We apply the Cartan's supersymmetric model to the electromagnetic and weak interaction of leptons and hadrons, and study dynamics of $\pi^0,\eta,\eta'$ meson and Higgs boson.  We adopt the Clifford algebra, express Pauli spinors by quaternions, Dirac spinors by octonions, and take into account the supersymmetric transformations $G_{ij}$ and $G_{ijk}$ of Cartan. 
analyzing the $B^0(\bar B^0)\to K^0(\bar K^0)\, J/\Psi$ decay processes.

When Higgs boson is assigned as Cartan's spinor pairs $\Psi=\bar\psi\psi$ or $\Phi=\bar\phi\phi$, its decay into $\gamma\gamma\gamma\gamma$ does not occur, but there appear 8 diagrams of $\Psi\to\gamma\gamma$ and 8 diagrams of $\Phi\to\gamma\gamma$.@

If our world matter makes 4.6\% of the total matter, and we can interpret the matter transformed by  $G_{23},G_{12},G_{123},G_{13}, G_{132}$ or about 5 times of our matter, i.e. 23\% corresponds to dark matter. 

We can assign the world transformed by $G_{12}, G_{123}, G_{13}$ and $G_{132}$, which make about 2/3 of the total corresponds to the dark energy. Experimentally, 72\% of the total matter corresponds to dark energy.

\end{abstract}

\newpage
\section{Introduction}
In the standard model, leptons and quarks are described by Dirac spinors, Majorana neutrinos are described by Pauli spinors, and proper description of fermion spinors is an important problem. 

In 2007, Georgi\cite{Ge07} proposed presence of unparticle stuff with a scale dimension $d_{\cal U}$ which looks like a nonintegral number  $d_{\cal U}$ of invisible particles. He discussed the possibility of a sector, which is scale-invariant and very weakly interacts with the rest of the standard world. The unparticle field that interact with particles of the standard model will give mass to these particles, which is scale invariant.  As the universe expands, the unparticle will produce Higgs boson\cite{Higgs64} which are 125Gev\cite{EWPD12}. Consequently, as the universe expands,  since the density of the Higgs is assumed to be invariant, the total mass of the universe will increase.  It violates the energy conservation rule. Georgi proposed that there could be the world which cannot be detected.

 Cartan\cite{Cartan66} formulated the coupling of 4-dimensional spinors $A,B,C,D$ and 4-dimensional vectors $E,E'$ using the Clifford algebra, which is a generalization of quaternions and octonions. 
In this model, there appears a triality symmetry, and one can construct sectors of $E$ and $E'$ which cannot be detected by fermions in the detector.  
Fermions in our universe are transformed by $G_{12},G_{13},G_{123}$ and $G_{132}$ to vectors, but the vectors produced by these transformations cannot be detected by electromagnetic probes in our universe. 

We applied the Cartan's supersymmetric model to the physical system and applied to the decay of $\pi^0, \eta, \eta'$ to $\gamma\gamma$ \cite{SF12a,SF12b,SF13a,SF13b}. The Pauli spinor was treated as a quaternion and the Dirac spinor was treated as an octonion. In the $\pi^0$ decay, the two final vector fields belong to the same group ($EE$) or ($E'E'$), and we call the diagram rescattering diagram.  In the decay of $\eta$, and $\eta'$, final vector fields belong to different groups ($EE'$), which we called twisted diagrams.  Qualitative difference of the $\eta$ decay or $\eta'$ decay, and $\pi^0$ decay can be explained by symmetry of Cartan's spinor.

We incorporate vector bosons and Higgs particles in the Cartan's supersymmetric model, and assign three physical states like $(e,\mu,\tau)$ leptons, $(\nu_1,\nu_2,\nu_3)$ neutrinos, etc. in definite triality states. 

In this paper we apply the model to $\pi^0,\eta$ or $\eta'$ decay into $\gamma\gamma$, to photoproduction of $\pi^0,\eta$ or $\eta'$, and study the physics of universe.

\section{Cartan's supersymmetric model and its extension to weak interaction}
Cartan defined matrix elements 
\begin{equation}
\mathcal F={^t\phi}{\mC X}\psi\nonumber
\end{equation}
where $^t\phi{\mathcal C}$ and $\psi$ are the final state and the initial state spinor, respectively and $X=\gamma_0 x^\mu \gamma_\mu$, i.e. electromagnetic interaction.

We first summarize the theory of  Cartan on the $SU(2)$ spinor.
He defined $\phi={^t(}A, B)$ with elements $\xi_{**}$ which
has an even number of indices and spinor $\psi={^t(}C,D)$ with elements $\xi_{*}$ which
has an odd number of elements.
\begin{eqnarray}
 \phi&=&A= \xi_{14}\sigma_x+\xi_{24}\sigma_y+\xi_{34}\sigma_z+\xi_{0} {\bf I},\nonumber\\
 \mC\phi&=&B= \xi_{23}\sigma_x+\xi_{31}\sigma_y+\xi_{12}\sigma_z+\xi_{1234} {\bf I},\nonumber\\
 \psi&=&C= \xi_{1}\sigma_x+\xi_{2}\sigma_y+\xi_{3}\sigma_z+\xi_{4} {\bf I},\nonumber\\
 \mC\psi&=&D= \xi_{234}\sigma_x+\xi_{314}\sigma_y+\xi_{124}\sigma_z+\xi_{123} {\bf I},
\end{eqnarray}
and added the vector fields expressed as
\begin{eqnarray}
&&E=x_1 {\bf i}+x_2{\bf j}+x_3{\bf k}+x_4{\bf I},\nonumber\\
&&E'={x_1}' {\bf i}+{x_2}'{\bf j}+{x_3}'{\bf k}+{x_4}'{\bf I}.
\end{eqnarray}

The spinors $A,B,C,D$ and the vectors $E,E'$ transform by superspace transformations,
$G_{23},G_{12},G_{13},G_{123}$ and $G_{132}$, and there is a triality symmetry among the
six spaces. 

\begin{table}
\leftline{Table 1. Transformations of $A,B,C,D,E$ and $E'$ by  $G_{23}$, $G_{12}$, $G_{123}$,$G_{13}$, $G_{132}$.}  
\begin{tabular}{lll}
$G_{23}$& $ G_{12}$& $G_{123}$\\
\hline
  $A\to (C_1,C_2,C_3,C_4)$ &  $A\to ({x_1}', {x_2}', {x_3}', x_4)$ &$A\to ({x_1}', {x_2}', {x_3}', {x_4}')$  \\
  $B\to (D_1,D_2,D_3,D_4)$ & $B\to (x_1,x_2,x_3,x_4')$ & $B\to (x_1,x_2,x_3,x_4)$   \\
  $C\to (A_1,A_2,A_3,A_4)$ & $C\to (C_1,C_2,C_3,D_4)$ & $C\to (A_1,A_2,A_3,B_4)$\\
  $D\to (B_1,B_2,B_3,B_4)$ & $D\to (D_1,D_2,D_3,C_4)$ &$D\to (B_1,B_2,B_3,A_4)$  \\
   $E\to (x_1,x_2,x_3,x_4')$ & $E\to (B_1,B_2,B_3,A_4)$ & 
 $E\to (D_1,D_2,D_3,C_4)$\\
    $E'\to (x_1',x_2', x_3', x_4)$ & $E'\to (A_1,A_2,A_3,B_4)$ &$E'\to (C_1,C_2,C_3,D_4)$ \\
\hline
\end{tabular}

\begin{tabular}{lll}
$G_{13}$& $G_{132}$& \\
\hline
$A\to (A_1,A_2,A_3,B_4)$ & $A\to (C_1,C_2,C_3,D_4)$&$A=(A_1,A_2,A_3,A_4)$ \\
$B\to (B_1,B_2,B_3,A_4)$&  $B\to (D_1,D_2,D_3,C_4)$&$B=(B_1,B_2,B_3,B_4)$ \\
  $C\to (x_1', x_2', x_3', x_4)$& $C\to (x_1',x_2',x_3',x_4)$&$C= (C_1,C_2,C_3,C_4)$ \\
 $D\to (x_1,x_2,x_3,x_4')$&  $D\to (x_1,x_2,x_3,x_4')$&$D= (D_1,D_2,D_3,D_4)$\\
 $E\to (D_1,D_2,D_3,D_4)$  & $E\to (B_1,B_2,B_3,B_4)$&$E= (x_1,x_2,x_3,x_4)$\\
 $E'\to (C_1,C_2,C_3,C_4)$ & $E'\to (A_1,A_2,A_3,A_4)$&$E'= (x_1',x_2', x_3', x_4')$ \\
\hline
\end{tabular}
\end{table}

The fermions $C,D$ produced by $G_{12}$ and $G_{132}$ are $(C_1,C_2,C_3,D_4)=\tilde C$, $(D_1,D_2,D_3,C_4)=\tilde D$ and $A,B$ produced by $G_{123}$ and $G_{13}$ are $(A_1,A_2,A_3,B_4)=\tilde A$, $(B_1,B_2,B_3,A_4)=\tilde B$ , i.e. the 4th component is interchanged. The fermions $A,B$  produced by $G_{23}$ are $(A_1,A_2,A_3,A_4)$, $(B_1,B_2,B_3,B_4)$ and $C,D$ produced by $G_{23}$ are $(C_1,C_2,C_3,C_4)$,$(D_1,D_2,D_3,D_4)$, i.e. the 4th component is not interchanged.  

The vector particles produced by $G_{12}, G_{13}$ and $G_{132}$ from fermions and produced by $G_{23}$ from $E,E'$ are $(x_1',x_2',x_3',x_4)=\tilde E'$, $(x_1,x_2,x_3,x_4')=\tilde E$, i.e. the 4th component is interchanged,  but that by $G_{123}$ is $(x_1',x_2',x_3',x_4')=E'$, $(x_1,x_2,x_3,x_4)=E$, i.e. the 4th component is not interchanged.  The difference of the 4th component is, however, the problem of complex conjugacy,  and not important for photons or Majorana neutrinos, which are treated as self-dual.
In order to incorporate the weak interaction, we consider
\[
{\mathcal G}={^t\phi} \mC X(1-\gamma_5)\psi=\sum_{i=1}^4[\overline{x^i}\langle\phi\mC \psi\rangle+\overline{{x^i}'}\langle \mC\phi \psi\rangle]
\]
where $\overline{x^i}$ or $\overline{{x^i}'}$ means that $-x^i\gamma_5$ or $-{x^i}'\gamma_5$  instead of $x^i$ or ${x^i}'$ part is chosen for $ ({^t\phi}X \mC \psi-{^t\phi} \mC X\psi)$ or $ ({^t\phi}X' \mC \psi-{^t\phi} X'\psi)$, respectively.

We observe that $\overline {x^4}(\mC{^t\phi}\mC \psi-{^t\phi}\mC\psi)$ which are chosen to be consistent with $x^i \mC\psi$ all contains $x^4\gamma_5 \mC \psi$ and $\overline {{x^4}'}({^t\phi}\mC \psi-{^t\phi}\psi)$ which are chosen to be consistent with ${x^i}'{^t\phi}$ all contain ${x^4}'{^t\phi}$. 
We assign $\phi$ mesons whose quark anti-quark pairs are expressed as $\xi_* \xi_{**}$, and the number of indices $*$ are both even, they are $\eta$ mesons, and $\psi$ mesons whose quark anti-quark pairs $\xi_* \xi_{**}$, and the number of indices $*$ are both odd, they are $\pi$ mesons. 

.  

In the analysis of $\pi^0,\eta$ and $\eta'$ decay into two photons, we treat quarks as four-component spinors, photons as selfdual vector particles, and Majorana neutrinos as selfdual spinors, and we consider operations of $G_{23}, G_{12},G_{13},G_{123}$ and $G_{132}$ on a set of fermions and vector particles.

We expect that the electromagnetic decays preserve triality, and the vector particles in a different triality sector cannot be detected by detectors containing leptons or quarks of different trialities, and they will appear as dark matter\cite{Xe12}. The two-component spinor belongs to a definite triality sector and photons, $W^\pm$ bosons, $Z$ bosons, Higgs boson are contained in the vector particles.

We choose ${^t(}E, E')$ or ${^t(}\tilde E, \tilde E')$ as a four component vector particle i.e. photon, $W^\pm$ boson, $Z$ boson or Higgs boson. We take into account a photon and a $Z$ boson decays into lepton pairs $\bar \ell \ell$ or neutrino pairs $\nu\bar\nu$ or $q\bar q$ and neglect more comlicated hadronic decays. A $W^\pm$ boson decays into $\ell\bar \nu$, $\bar \ell\nu$ or $q\bar q$.

\subsection{The transformation by $G_{23}$}
The electromagnetic transition of a lepton or a quark ${^t(}A,B)$ to ${^t(}D,C)$ occurs 
by the operator $G_{23}$ as
\[
\left(\begin{array}{c}A\\
                          B\end{array}\right)\to\left(\begin{array}{c}{ \tilde E} \\
                                                                                {\tilde E'}\end{array}\right)\left(\begin{array}{c}D\\
                          C\end{array}\right)=\begin{array}{c}{\gamma, Z}\\
                                                                      {\ell\bar\ell, q\bar q, \nu\bar\nu}\end{array} \left(\begin{array}{c}D\\
                          C\end{array}\right),
\]
\[
\left(\begin{array}{c}D\\
                          C\end{array}\right)\to\left(\begin{array}{c}{ \tilde E'}\\
                                                                                 { \tilde E}\end{array}\right)
\left(\begin{array}{c}A\\
                          B\end{array}\right)=\begin{array}{c}{\gamma, Z}\\
                                                             {\ell\bar \ell, q\bar q,\nu\bar\nu}\end{array}\left(\begin{array}{c}A\\
                          B\end{array}\right).
\]

In the theory of Glashow-Weinberg-Salam theory\cite{PS95,BBJ81}, in addition to the photon current contribution, which we denote by $\gamma$, there is the $Z$ boson current and $W$ boson current contribution in the interaction of quarks and leptons.  

The operator $G_{23}$ transforms electromagnetic field $E$ to $\tilde E$ and $E'$ to $\tilde E'$, or in the wave function, transformation  $e^{iE t}$ to $e^{i\tilde Et}$ or $e^{iE't}$ to $e^{i\tilde E't}$ is equivalent to the time reversal operation. If the gravitational force is insensitive to the electromagnetic triality selection rule, the regions sensitive to all $G_{23},G_{12},G_{123}$, $G_{13},G_{132}$ interactions will be five times larger than the region sensitive to $G_{23}$.
Anti leptons and self-dual gluons are produced by an operation of $G_{23}$, and the triality sectors are preserved.

\subsection{The transformation by $G_{13}$}
The operation $G_{13}$ on the lepton or quark ${^t(}A,B)$ produces the spinor whose direction of time is same as the original state${^t(} \tilde B,\tilde A)$, and on the lepton or quark ${^t(}C,D)$ produces a vector boson ${^t(} E' ,E)$, which can be identified as $W^{\pm}$ which decays to $\ell\bar \nu$, ${\bar\ell}\nu$ or $q\bar q$. We assume that the lepton $l$ produced from the vector field $E$ and the lepton $l'$ produced from the vector field $E'$ pair annihilate.

\[
\left(\begin{array}{c}A\\
                          B\end{array}\right)\to
\left(\begin{array}{c}{\tilde B}\\
                          {\tilde A}\end{array}\right)
\left(\begin{array}{c}{l\tilde E'}\\
                          {l'\tilde E} \end{array}\right) \to
 \left(\begin{array}{c}{\tilde B}\\
                          {\tilde A}\end{array}\right)
\begin{array}{c}{W^\pm}\\
                   {\ell\bar\nu, \bar \ell\nu, q\bar q}\end{array} .                 
\]

\subsection{The transformation by $G_{132}$}
The operation of $G_{132}$ on a lepton or a quark ${^t(}A,B)$ produces the spinor whose direction of time is same as the original state ${^t(}\tilde D,\tilde C)$, and that on a quark ${^t(}C,D)$ produces a vector boson  ${^t(}\tilde E', \tilde E)$.  We assume that the lepton $l$ produced from the vector field $E$ and the lepton $l'$ produced from the vector field $E'$ pair annihilate.

\[ 
\left(\begin{array}{c}A\\
                          B\end{array}\right)\to
\left(\begin{array}{c}{\tilde D}\\
                          {\tilde C}\end{array}\right)
\left(\begin{array}{c}l{\tilde E'}\\
                          {l'\tilde E} \end{array}\right)  \to
 \left(\begin{array}{c}{\tilde D}\\
                          {\tilde C}\end{array}\right)
 \begin{array}{c}{W^{\pm}} \\
                        {\ell\bar \nu, \bar\ell \nu, q\bar q}\end{array}.              
\]

\subsection{The transformation by $G_{12}$}
The operator $G_{12}$ transforms the spinor field ${^t(} A,B)$ to the vector field ${^t(} \tilde E',\tilde E)$, and a lepton or a quark state ${^t(}C,D)$ to the spinor whose direction of time is same as the original state${^t(} \tilde D,\tilde C)$. 
We assume that the lepton $l$ produced from the vector field $E$ and the lepton $l'$ produced from the vector field $E'$ pair annihilate..

\[
\left(\begin{array}{c}C\\
                          D\end{array}\right)\to
\left(\begin{array}{c}{\tilde D}\\
                          {\tilde C}\end{array}\right)
\left(\begin{array}{c}{l \tilde E'} \\
                         {l' \tilde E} \end{array}\right)\to
\left(\begin{array}{c}{\tilde D}\\
                          {\tilde C}\end{array}\right) 
\begin{array}{c} {W^{\pm}}\\
                     {\ell\bar\nu, \bar\ell\nu, q\bar q}\end{array}.               
\]

\subsection{The transformation by $G_{123}$}
Operations of $G_{123}$ on  a lepton or a quark state ${^t(} C,D)$ produces an anti-lepton or an anti-quark state ${^t(}\tilde B, \tilde A)$, and on a lepton or a quark state ${^t(}A,B)$ produces a vector state ${^t(} E',E)$.
We assgn the vector field $E(E')$ as $W^\pm$ which can decay into $\ell\bar \nu$, $\bar\ell \nu$, $q\bar q$.

\[
\left(\begin{array}{c}C\\
                          D\end{array}\right)\to
\left(\begin{array}{c}{\tilde B}\\
                          {\tilde A}\end{array}\right)
\left(\begin{array}{c}{l E' }\\
                          {l' E  }\end{array}\right)\to
 \left(\begin{array}{c}{\tilde B}\\
                          {\tilde A}\end{array}\right) 
 \begin{array}{c}  {W^\pm}\\
                      {\ell\bar\nu, \bar\ell\nu, q\bar q}\end{array}.              
\]

\subsection{Application to the decay of $B^0(\bar B^0)\to K^0_L(\bar{K^0_L})\, J/\Psi$}
The transformations $G_{**}$ and $G_{***}$ can be checked by studying decays of $B^0
(\bar B^0)$ to $K^0_L(\bar{K^0_L})\, J/\Psi$\cite{SF15}.

\begin{figure}
\begin{minipage}[b]{0.47\linewidth}
\begin{center} 
\includegraphics[width=4cm,angle=0,clip]{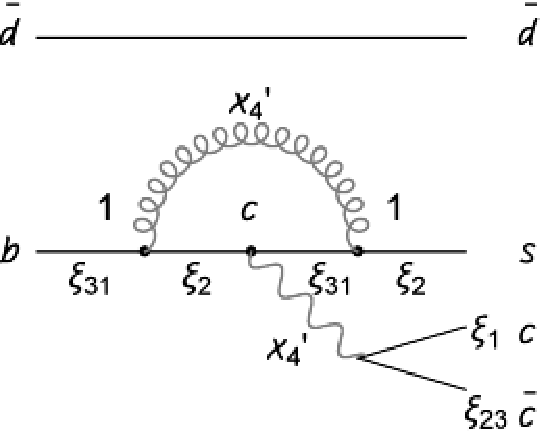} 
\end{center}
\caption{Typical diagrams of  $B^0\to K_L^0\, J/\Psi$ decay. Penguin diagram $1\,1$ type . In the decay channel of $b=\xi_{31}$, ${x^4}'$ decays to $\xi_{1}\xi_{23}$ or $\xi_{3}\xi_{12}$.}
\label{bscc11a}
\end{minipage}
\hfill
\begin{minipage}[b]{0.47\linewidth}
\begin{center}
\includegraphics[width=4cm,angle=0,clip]{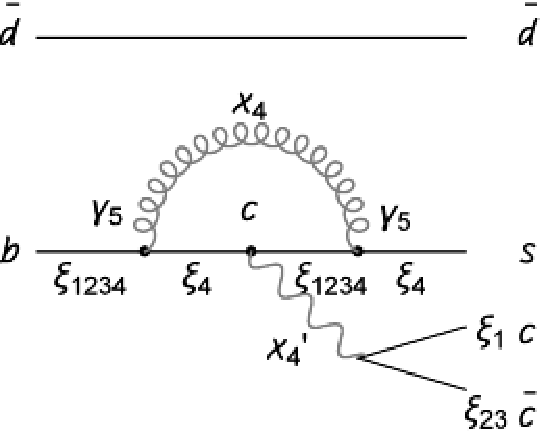} 
\end{center}
\caption{Typical diagrams of  $B^0\to K_L^0\, J/\Psi$ decay. Penguin diagram $\gamma_5\gamma_5$. In the decay channel of $b=\xi_{1234}$, ${x^4}'$ decays to $\xi_{1}\xi_{23}$ or $\xi_{2}\xi_{31}$ or $\xi_{3}\xi_{12}$. }
\label{bscc11b}
\end{minipage}
\end{figure}

The decay of $B^0\to K_L^0\, J/\Psi$ is explained by the penguin diagram which is shown in Figs.\ref{bscc11a} and \ref{bscc11b}. Emission of a $W$ boson from $b$ quark $\xi_{31}$ or $\xi_{1234}$ is done by the operation of $G_{13}$, and absorption to $s$ quark $\xi_2$ or $\xi_4$ is done by $G_{132}$. Emission of a photon $x_4'$ from the quark $\xi_2$ or $\xi_4$ and changing to 
the quark $\xi_{123}$ or $\xi_{1234}$ is performed by $G_{23}$

\begin{figure}
\begin{minipage}[b]{0.47\linewidth}
\begin{center} 
\includegraphics[width=4cm,angle=0,clip]{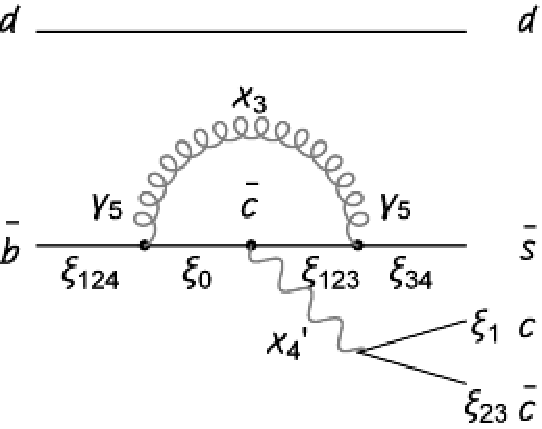}  
\end{center}
\caption{Typical diagrams of  $\bar B^0\to K_L^0\, J/\Psi$ decay. Penguin diagrams $\gamma_5\gamma_5$ type. In the diagrams, ${x^4}'$ decays to  $\xi_{1}\xi_{23}$ or $\xi_{2}\xi_{31}$ or $\xi_{3}\xi_{12}$. }
\label{bscc55a}
\end{minipage}
\hfill
\begin{minipage}[b]{0.47\linewidth}
\begin{center}
\includegraphics[width=4cm,angle=0,clip]{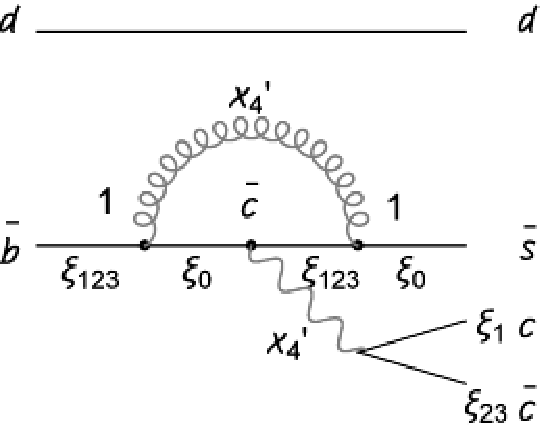}
\end{center}
\caption{Typical diagrams of  $\bar B^0\to K_L^0\, J/\Psi$ decay. Penguin diagrams $\gamma_5\gamma_5$ type. In the diagrams, ${x^4}'$ decays to  $\xi_{1}\xi_{23}$ or $\xi_{2}\xi_{31}$ or $\xi_{3}\xi_{12}$. }
\label{bscc55b}
\end{minipage}
\end{figure}

The decay of $\bar B^0\to \bar K_L^0\, J/\Psi$ is shown in Figs.\ref{bscc55a} and\ref{bscc55b}. Emission of a $W$ boson from $\bar b$ quark is done by the operation of $G_{123}$ and absorption to $s$ quark is done by the operation of $G_{132}$. Emission of a photon is done by the operation of $G_{23}$.

\begin{figure}
\begin{minipage}[b]{0.47\linewidth}
\begin{center}
\includegraphics[width=4cm,angle=0,clip]{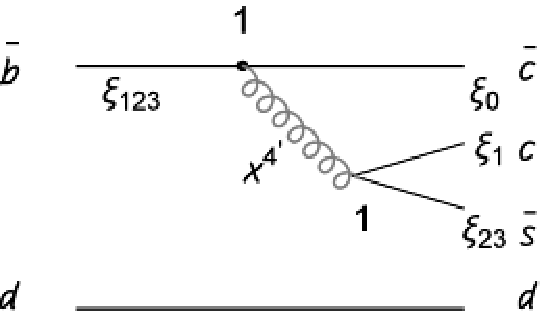}  
\end{center}
\caption{Tree diagrams of  $\bar B^0\to \bar K_L^0\, J/\Psi$ decay $\gamma_5\gamma_5$ type. In the diagrams, $x_4'$ decays to $\xi_{3}\xi_{24}$. }
\label{bbarT5523a}
\end{minipage}
\hfill
\begin{minipage}[b]{0.47\linewidth}
\begin{center} 
\includegraphics[width=4cm,angle=0,clip]{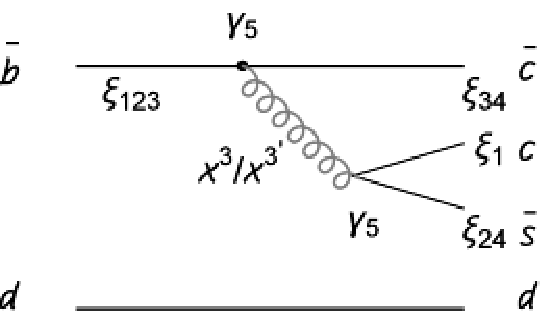} 
\end{center}
\caption{Tree diagrams of  $\bar B^0\to \bar K_L^0\, J/\Psi$ decay $\gamma_5\gamma_5$ type. In the diagrams, $x^3/{x^3}'$ decays to $\xi_{1}\xi_{24}$. }
\label{bbarT5523b}
\end{minipage}
\end{figure}

In the case of decay from $\bar B^0$, there are tree diagrams, in which we allow the vector field of $E(E')$ type is created and as it propagate in the vacuum it changes to $E'(E)$ type and create $q\bar q$ meson, as shown in Fig,\ref{bbarT5523b}.

The BABAR  collaboration\cite{BaBar09} measured the CP asymmetry of the flavor tagged $\Delta t$ distribution of
\[
A_{CP,f}=\frac{\Gamma_{\bar B^0\to f}(\Delta t)-\Gamma_{B^0\to f}(\Delta t)}
{\Gamma_{\bar B^0\to f}(\Delta t)+\Gamma_{B^0\to f}(\Delta t)}
\]
and in the $\bar B^0$ tagged events, observed enhancement around about $|\Delta t|=5$ps.
To check whether it is due to the tree diagrams of  $\bar B^0\to \bar K_L^0\, J/\Psi$, it is necessary to calculate the propagator of $W$ boson in detail, and it is reft to the future.

The creation of $c\bar c$ in penguin diagram and in tree diagram should be different. Bitbol\cite{Bitbol85} discussed time asymmetry that could emerge when two measurements are
performed by single observer but in opposite direction of time. The direction of motion of the $W$ boson could also cause the time reversal asymmetry, and the CP asymmetry. The problem is under investigation.
\section{Quaternions in $\pi^0,\eta$ or $\eta'$ decay into $\gamma\gamma$}
In the world of elementary particles, there are fermions which are transformed by quaternions and expressed by two-component spinors or, which are transformed by octonions and expressed by four component spinors. 

 In Quantum Chromo Dynamics( QCD ), complex numbers and quaternions are used.  A quaternion operates on a
 two-component spinor, i.e.  Pauli spinor.  The Dirac spinor is a four component spinor, but the octonion transforms as a four component spinor, which has the triality symmetry.   

Phenomenologically, decay of a pion into two gamma rays is well described
by a divergence of the axial current, since a pion can be regarded as a Nambu-Goldstone boson\cite{Adler65}.
The Adler-Bardeen's theorem\cite{AB69} says that higher-order effects in the triangular diagram
can be incorporated in the renormalization, represented by a rescattering diagram.

The theoretical decay width of $\pi^0$ and that of $\eta$ into two $\gamma$ are
\[
\Gamma(\pi^0\to\gamma\gamma)=\frac{\alpha^2}{32\pi^3}\frac{{m_\pi}^3}{{f_\pi}^2}=7.7{\rm eV},
\]
and
\[
\Gamma(\eta\to\gamma\gamma)=\frac{\alpha^2}{32\pi^3}\frac{1}{3}\frac{{m_\eta}^3}{{f_\eta}^2}=0.13{\rm keV},
\]
where $f_\pi=137$MeV and $f_\eta=150$MeV.

\begin{figure}[htb]
\begin{minipage}[b]{0.47\linewidth}
\begin{center}
\includegraphics[width=4cm,angle=0,clip]{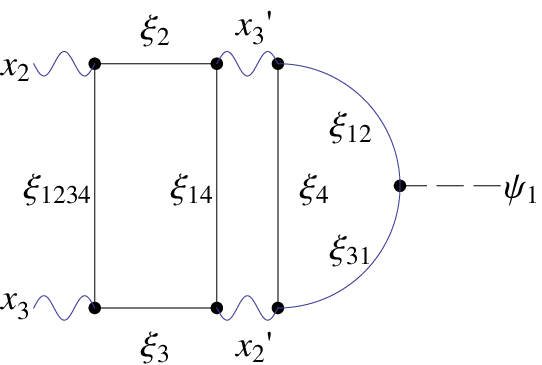}%
\caption{The half circle diagram of axial anomaly. $\psi_1{x_3}' {x_2}'$ isovector type and its rescattering.} 
\label{g_psi1}
\end{center}
\end{minipage}
\hfill
\begin{minipage}[b]{0.47\linewidth}
\begin{center}
\includegraphics[width=4cm,angle=0,clip]{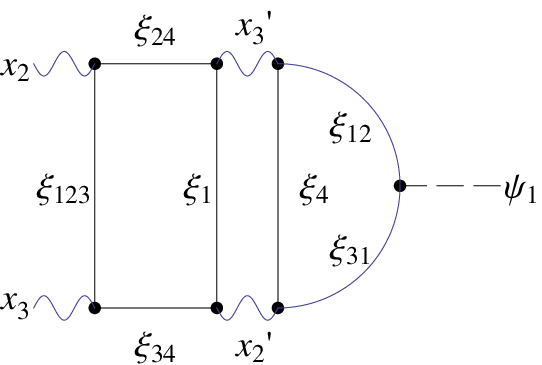}%
\caption{The half circle diagram of axial anomaly. $\psi_1x_3 x_2$ isovector type and its rescattering.} 
\label{g_psi2}
\end{center}
\end{minipage}
\begin{minipage}[b]{0.47\linewidth}
\begin{center}
\includegraphics[width=4cm,angle=0,clip]{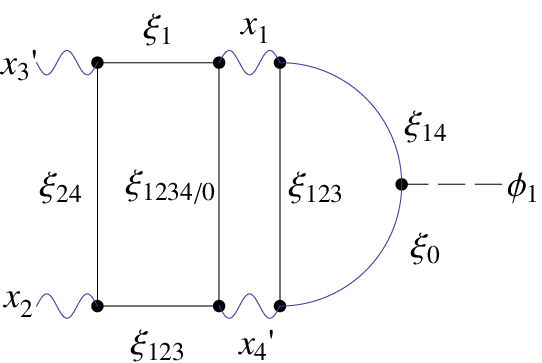}
\caption{The half circle diagrams of $\eta(\eta')$ decay. $\phi_1,x_1,{x_4}'$ type and its rescattering.	.}
\label{g_phi1}
\end{center}
\end{minipage}
\hfill
\begin{minipage}[b]{0.47\linewidth}
\begin{center}
\includegraphics[width=4cm,angle=0,clip]{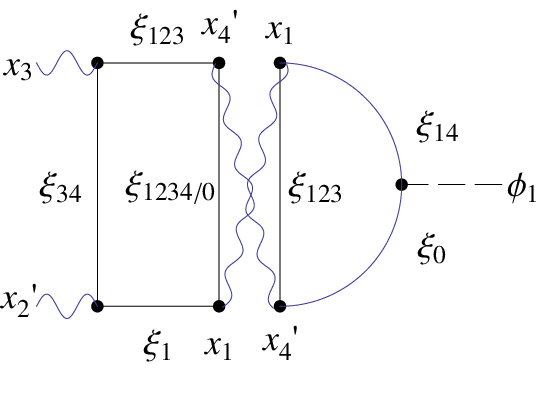}
\caption{The half circle diagrams of $\eta(\eta')$ decay. $\phi_1,x_1,{x_4}'$ type and its rescattering.. }
\label{g_phi2}
\end{center}
\end{minipage}
\end{figure}

 Experimental decay width of $\pi^0$ is consistent with the theory, but that of $\eta$ is about 4 times larger.
The decay of $\pi^0$ can be estimated from the half-circle diagram with two total spin 1, relative p-wave isovector photons in final states, and its diagram followed by rescattering diagrams as shown in Figs.\ref{g_psi1} and \ref{g_psi2}.
In the case of decay of $\eta$ or $\eta'$, the two total spin 1, relative p-wave isoscalar photons, and its diagram followed by rescattering is expected. When the two photons are the same type e.g. $x_2,x_3$ or ${x_2}',{x_3}'$ and relative p-wave, it will be impossible to make total isospin equals 0. 

Therefore, we choose two photons of different types e.g. $x_2, {x_3}'$ or ${x_2}',x_3$ and consider the rescattering, as shown in Fig.\ref{g_phi1} and \ref{g_phi2}.  Inside hadrons, as the Table.1 shows, the operators $G_{123}$ and $G_{13}$ interchange the quark $A_4=\xi_0$ and the quark $B_4=\xi_{1234}$,  and the operators $G_{132}$ and $G_{12}$ interchange the quark $C_4=\xi_4$ and $D_4=\xi_{123}$.  When $\xi_{123}$ absorbs $x_4'$, whether it is transformed to $\xi_{1234}$ or to $\xi_0$ will not be crucial. In Fig. \ref{g_phi1}, $x_1,{x_3}',x_2,{x_4}'$ photons or gluons contribute and in Fig.\ref{g_phi2} $x_1, {x_2}',x_3,{x_4}'$ photons or gluons contribute.  The contribution of Fig.\ref{g_phi1} and Fig.\ref{g_phi2} are coherent and the amplitude of $\eta$ or $\eta'$ decay into two photons will become 4 times larger than the incoherent sum.

 Two $\gamma$'s in $\pi^0$ decay belong to the same sector of vector particles $(E,E)$ or                                                    
$(E', E')$, but two $\gamma$'s from $\eta (\eta')$ can be
 $(E, E')$ or  $(E', E)$ from twisted diagram contribution, which could enhance the decay 
width by about factor five\cite{SF13a,SF13b}. 
Experimentally\cite{CBTA09}, decay of $\eta$ and $\eta'$ are analyzed via $\eta\to 3\pi^0\to 6\gamma$ and $\eta'\to \pi^0\pi^0\eta\to 6\gamma$.

In the $\eta(\eta')$ decay into two photons, superposition of a propagation of $\xi_{1234}$ and $\xi_0$ 
was incorporated in the decay into $x_3'$ and $x_2$. 
\begin{figure}[htb]
\begin{minipage}[b]{0.47\linewidth}
\begin{center}
\includegraphics[width=4cm,angle=0,clip]{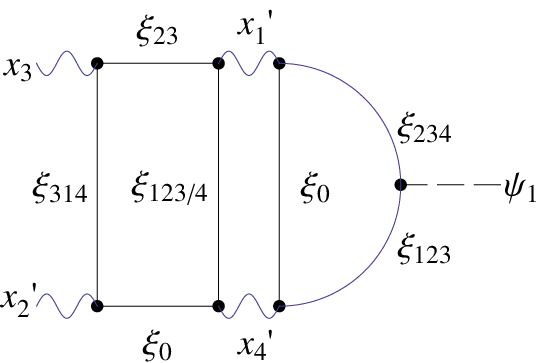}
\caption{The improper half circle diagrams of $\eta(\eta')$ decay. $\phi_1,{x_1}',{x_4}'$ type and its rescattering. }
\label{gb1e}
\end{center}
\end{minipage}
\hfill
\begin{minipage}[b]{0.47\linewidth}
\begin{center}
\includegraphics[width=4cm,angle=0,clip]{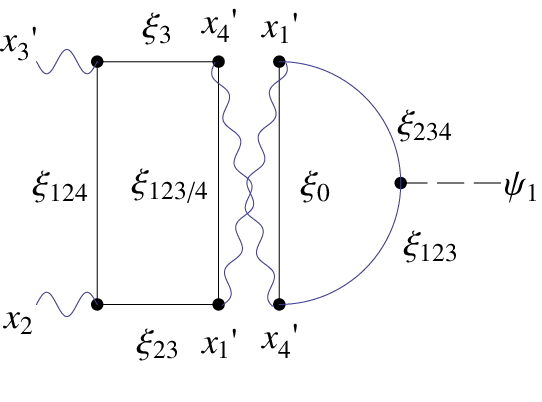}
\caption{The improper half circle diagrams of $\eta(\eta')$ decay. $\phi_1,{x_1}',{x_4}'$ type and its rescattering. }
\label{gb1ft}
\end{center}
\end{minipage}
\end{figure}
They are coupled to the intermediate ${x_1}',{x_4}'$ state, which is a $\pi$ meson. Therefore the diagrams 
of Figs.\ref{gb1e} and \ref{gb1ft} are improper. 

Two photon decays of $\pi^0,\eta$ and $\eta'$ and $\eta,\eta'$ mixing were studied by \cite{DHL85, BBC88, KaLe00, BoWe01, BeBo01,BoNi03,Shore01}. The $\eta, \eta'$ mixing from Lattice simulation was studied in\cite{MOU13}.

\section{Quaternions in $\pi^0, \eta$ or $\eta'$ photoproduction near threshold}
Near threshold, pseudoscalar meson photoproduction was studied in chiral perturbation theory to one loop  in \cite{GL84}.  In this theory, effective current quark mass
\[
\hat m=\frac{1}{2}(m_u+m_d) 
\]
of the order of 7MeV is incorpolated and with characteristic scale $M\sim 500-1000$MeV, perturbative correction of the order of $\hat m/M$ is considered.  

\begin{figure}[htb]
\begin{minipage}[b]{0.47\linewidth}
\begin{center}
\includegraphics[width=4cm,angle=0,clip]{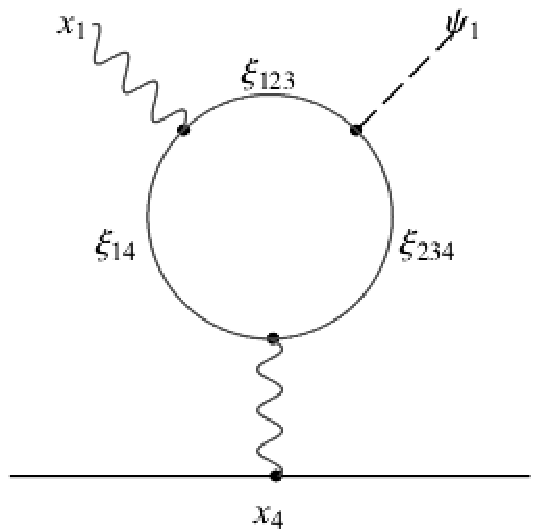}
\caption{Polarization $+1$ photon $\pi^0$ production vertex.  $\psi_1 x_1 x_4$ type. }
\label{Ps114123}
\end{center}
\end{minipage}
\hfill
\begin{minipage}[b]{0.47\linewidth}
\begin{center}
\includegraphics[width=4cm,angle=0,clip]{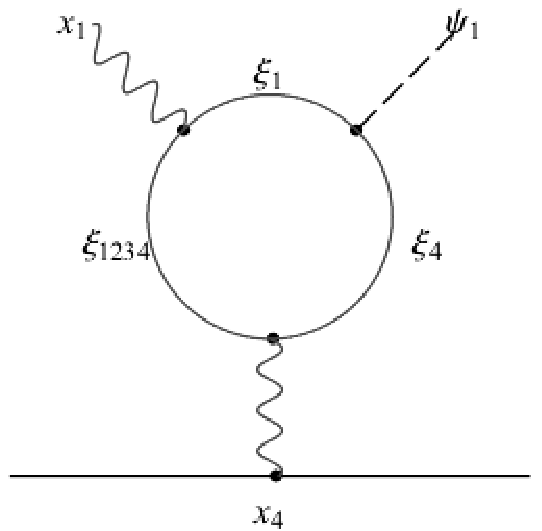}
\caption{Polarization $+1$ photon $\pi^0$ production vertex.  $\psi_1 x_1 x_4$ type. }
\label{Ps1141}
\end{center}
\end{minipage}
\begin{minipage}[b]{0.47\linewidth}
\begin{center}
\includegraphics[width=4cm,angle=0,clip]{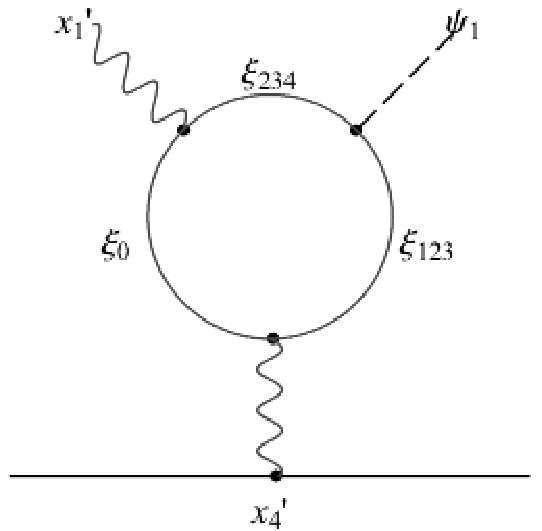}
\caption{Polarization $+1$ photon $\pi^0$ production vertex. $\psi_1 {x_1}' {x_4}'$ type. }
\label{Ps11P4P234}
\end{center}
\end{minipage}
\hfill
\begin{minipage}[b]{0.47\linewidth}
\begin{center}
\includegraphics[width=4cm,angle=0,clip]{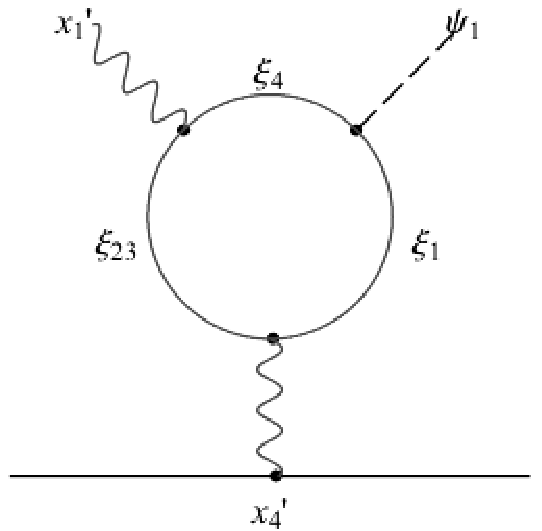}
\caption{Polarization +1\, photon $\pi^0$ production vertex.  $\psi_1 {x_1}' {x_4}'$ type. }
\label{Ps11P4P4}
\end{center}
\end{minipage}
\end{figure}

Near threshold, we consider diagrams of the type Figs.\ref{Ps114123}-\ref{Ps11P4P4}, i.e. exchange of photons, which may effectively appear as a meson production in a cloud of electron-positron pairs. 
In the pion photoproduction vertex, the types of the incoming photon and the exchanged photon are assumed to belong to the same triality sector.

Photoproduction of $\eta$ or $\eta'$ is not straightforward to analyze, since the pion photoproduction followed by the final state interaction of $\pi N\to\eta N$ or $\eta' N$ are not easy to estimate\cite{SHKM10}.

We can, however, speculate qualitative difference of  $\eta$ or $\eta'$ meson photoproduction vertex from pion photoproduction vertex near threshold.
In the case of $\eta$ or $\eta'$ photoproduction, the types of the incoming photon and the exchanged photon need to belong to different types.   

The spinor of the quark and anti quark pair can have polarization +1 or -1.  Combined with the photon
polarization, total angular momentum becomes 0.

In \cite{SF13b}, we found that the s-wave mesons, whose Cartan's quark anti quark pair have both even number of indices or both odd number of indices, the masses become light. In the pseudoscalar meson photoproduction near threshold, we consider a quark pair creation near a proton in the background of Coulomb field. We regard in this section,  $\psi_*$ meson as a $\pi^0$ meson.
The PCAC hypothesis constraints the current as\cite{Hab00}
\[
(p'-p)_\mu j^\mu_A=-\frac{f_\pi \mu^2}{q^2-\mu^2}\bar u_f \gamma_5 G_t u_i \tau
\]
and \cite{Adler65} finds the amplitude of photoproduction as
\[
\frac{f_\pi \mu^2}{q^2-\mu^2}{\mathcal M}=q_\mu J^{\mu\nu}_{A,\gamma}\epsilon_\nu-Q_\pi j^\nu_A \epsilon_\nu.
\]

Here, $J^{\mu\nu}_{A,\gamma}$ is the axial current that the photon couples,  $j^\nu_A$ is the 
nucleon matrix element, $\tau$ is the vertex isospin operator, and $(Q_\pi)_{kl}=e\, i\, \epsilon_{k3l}$ is the pion charge operator.

In the soft pion limit $q\to 0$, the amplitude becomes
\[
{\mathcal M}|_{q=0}=-e_\pi \bar u_f\frac{\gamma_5 \gamma^\nu}{2m} \tilde G_A(k^2)u_i\epsilon_\nu,
\]
where $e_\pi=Q_\pi\tau$. 
 
Contributions of the polarization $+1$ photon in $\pi^0$ production on a proton at the threshold are presented in Figs. \ref{Ps114123}-\ref{Ps11P4P4}. 

\begin{figure}[htb]
\begin{minipage}[b]{0.47\linewidth}
\begin{center}
\includegraphics[width=4cm,angle=0,clip]{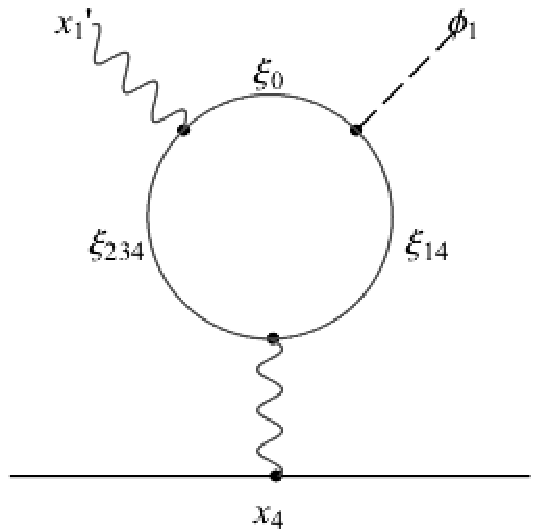}%
\caption{Polarization $+1$ photon $\eta$ production vertex. $\phi_1{x_1}' x_4$type.} 
\label{Pi1P40}
\end{center}
\end{minipage}
\hfill
\begin{minipage}[b]{0.47\linewidth}
\begin{center}
\includegraphics[width=4cm,angle=0,clip]{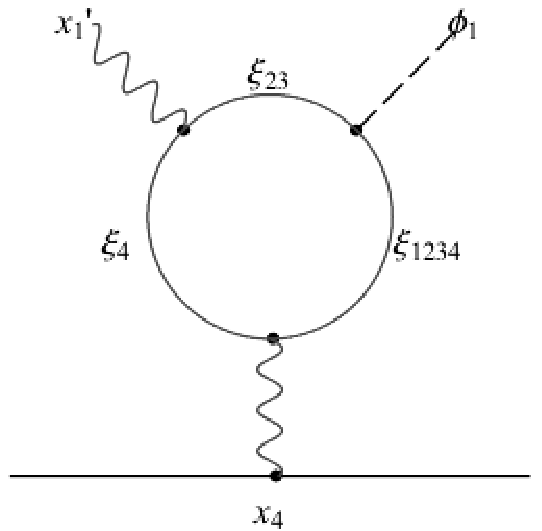}
\caption{Polarization $+1$ photon $\eta$ production vertex. $\phi_1 {x_1}' x_4$ type.}
\label{Pi1P423}
\end{center}
\end{minipage}
\begin{minipage}[b]{0.47\linewidth}
\begin{center}
\includegraphics[width=4cm,angle=0,clip]{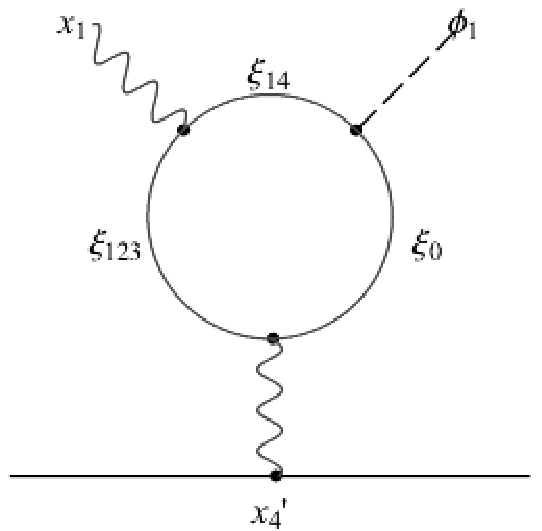}%
\caption{Polarization $+1$ photon $\eta$ production vertex. $\phi_1x_1 {x_4}'$type.} 
\label{Pi14P14}
\end{center}
\end{minipage}
\hfill
\begin{minipage}[b]{0.47\linewidth}
\begin{center}
\includegraphics[width=4cm,angle=0,clip]{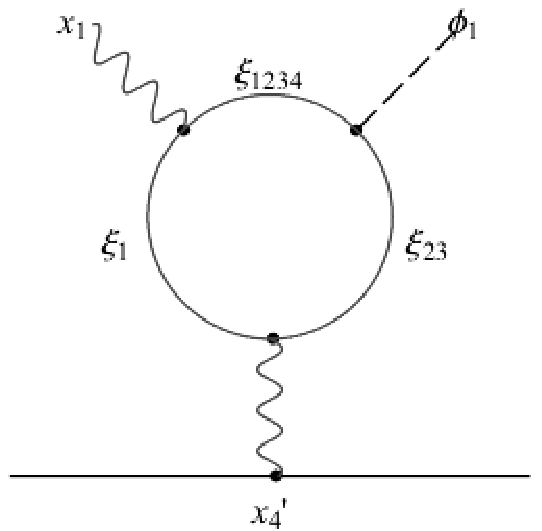}
\caption{Polarization $+1$ photon $\eta$ production vertex. $\phi_1 x_1 {x_4}'$ type.}
\label{Pi1P1234}
\end{center}
\end{minipage}
\end{figure}

Contributions of the photons of polarization $+1$ in $\eta$ production on a proton at the threshold are presented in Figs. \ref{Pi1P40}-\ref{Pi1P1234}. 

In the meson photoproduction, contribution of photons of the polarization $-1$ is similar,
Although $\pi$ mesons are mainly restricted in the $u-d$ sector, mixing of the $u-d$ sector and the $c-s$ sector in $\eta$ and $\eta'$ mesons is important, and a detailed analysis remains as a future problem.

\newpage
\section{Decays of a Higgs boson into photons}
If a scalar boson consists of Cartan's spinor pairs $\Psi=\bar\psi \psi$ or $\Phi=\bar\phi\phi$, it cannot directly decay into two photons, since the coupling of the spinor and the  
 photon is restricted to the $^t\phi CX\,\psi$ type, and the decay into 4 photons is the simplest.

Experimentally the fraction of Higgs boson decay into 4 photons is small, and decays of Higgs bosons into 4 leptons and subsequent pair annihilations into two photons are observed\cite{CMS11,ATLAS12,ATLAS11,CMS13,ATLAS14a,ATLAS14b}. 
Visible cascade Higgs decays to four photons via a CP odd state: $2a$ \cite{DLM00},  or via two photons and two gluons state: $2g2\gamma$ were discussed \cite{CFW07}.   

\subsection{Higgs boson decay into $\gamma\gamma\gamma\gamma$}
Before studying the decay into 4 leptons, we study the decay of a scalar boson expressed by Cartan's spinor pair  into 4 photons.  There are two types of Cartan's spinor which are relatively anti-particles, $\phi\bar\phi$ and $\psi\bar\psi$.

The scalar boson $\Phi(\phi\bar\phi)$ decays into quark pairs  $\phi \mC\phi$, (or $\mC\phi \phi$) and the scalar boson $\Psi(\psi\bar\psi)$ decays into quark pairs $\psi \mC\psi$ (or $\mC\psi \psi$). The quarks produce photons, and we choose the direction of the momenta of produced photons $E$ or $E'$, along $x_1, x_3$,  $x_1, x_3'$, ${x_1}',x_3$ or ${x_1}', {x_3}'$ and  subsequent directions of the photons along ${x_1}',{x_3}'$, ${x_1}',x_3$, $x_1,{x_3}'$ or $x_1,x_3$, respectively. 

If a scalar boson $\Psi(11)$ decays into lepton pairs $\xi_4 \mC\xi_4$ (or $\mC\xi_4 \xi_4$), and after the emission of two photons $x_3'$ and $x_1'$, lepton pairs $\xi_{12}, \mC\xi_{23}$ (or $\mC\xi_{12}, \xi_{23})$ are produced, which emit secondary photons $x_1,x_3$, and the secondary photons  are joined by the lepton $\xi_{314}$ (or $\mC\xi_{314}$), we specify the process as
\[
\Psi(11)\to \mC\phi \mC\phi, \mC\psi \quad (or\quad \Psi(11)\to \phi \phi, \psi)
\] 
since 
$\xi_{12}$ and $\xi_{23}$ belong to the $\mC\phi$ group and $\xi_{314}$ belongs to the $\mC\psi$ group, but the
choice of the basis $\xi_4 \mC\xi_4$ or $\mC\xi_4 \xi_4$ is arbitrary. There are 8 diagrams of $\Psi$ decay into 4 photons,  and 8 diagrams of $\Phi$ decay into 4 photons, when directions of the momenta of photons are chosen to be along the $x_1$ direction or the $x_3$ direction.

The decay amplitude of $\Psi(\psi\bar\psi)$ has the corresponding decay amplitude of $\Phi(\phi\bar\phi)$ and the amplitude of $x^i({x^i}')\xi_*\xi_{**}$ from $\Psi$ decay and from $\Phi$ decay have opposite phase, and we expect the scalar boson decay into  $\gamma\gamma\gamma\gamma$ does not occur.
 
\begin{figure}[htb] 
\begin{minipage}[b]{0.47\linewidth}
\begin{center}
\includegraphics[width=4cm,angle=0,clip]{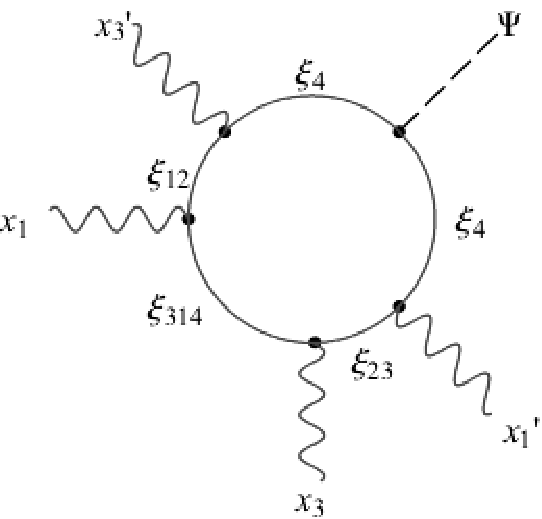}
\caption{Scalar boson decay into 4 photons.  $\Psi (11)\to \mC\phi \mC\phi,\mC\psi$ type. }
\label{scalarPsi314}
\end{center}
\end{minipage}
\hfill
\begin{minipage}[b]{0.47\linewidth}
\begin{center}
\includegraphics[width=4cm,angle=0,clip]{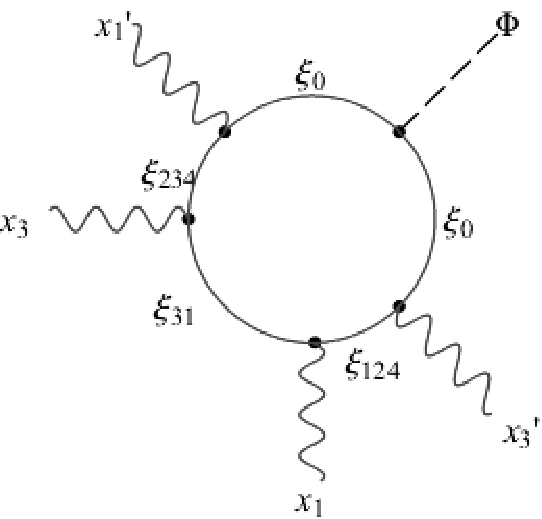}
\caption{Scalar boson decay into 4 photons. $\Phi (11)\to \mC\psi \mC\psi, \mC\phi$ type. }
\label{scalarPhi31 }
\end{center}
\end{minipage}
\begin{minipage}[b]{0.47\linewidth}
\begin{center}
\includegraphics[width=4cm,angle=0,clip]{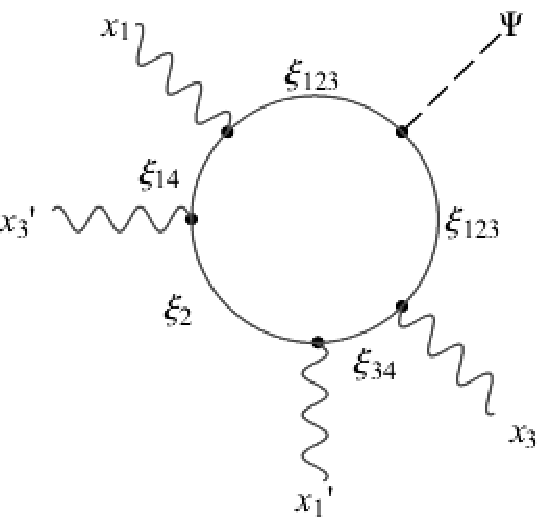}
\caption{ $\Psi (11)\to \phi \phi,\psi$ type.} 
\label{scalarPsi234}
\end{center}
\end{minipage}
\hfill
\begin{minipage}[b]{0.47\linewidth}
\begin{center}
\includegraphics[width=4cm,angle=0,clip]{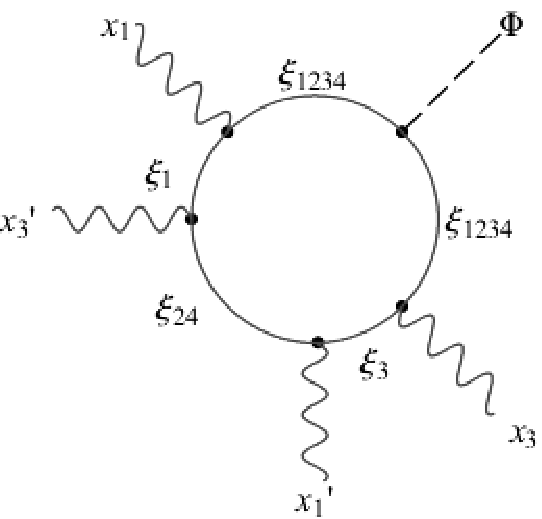}
\caption{ $\Phi (11)\to\psi\psi,\phi$ type.} 
\label{scalarPhi4a}
\end{center}
\end{minipage}
\end{figure}
\begin{figure}[htb]
\begin{minipage}[b]{0.47\linewidth}
\begin{center}
\includegraphics[width=4cm,angle=0,clip]{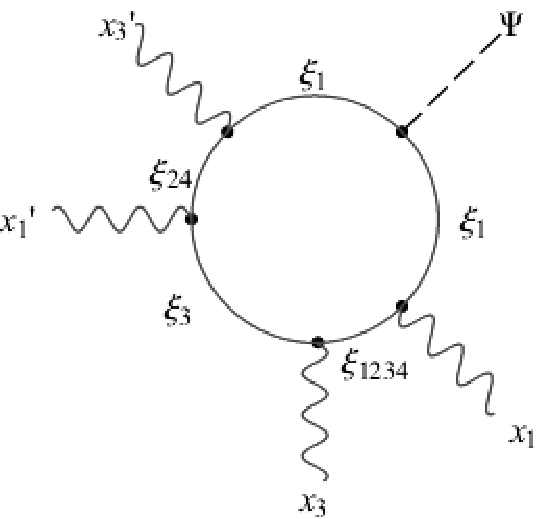}
\caption{ $\Psi (\Vec{i i})\to\phi \mC\phi, \psi$ type.} 
\label{scalarPsi3}
\end{center}
\end{minipage}
\hfill
\begin{minipage}[b]{0.47\linewidth}
\begin{center}
\includegraphics[width=4cm,angle=0,clip]{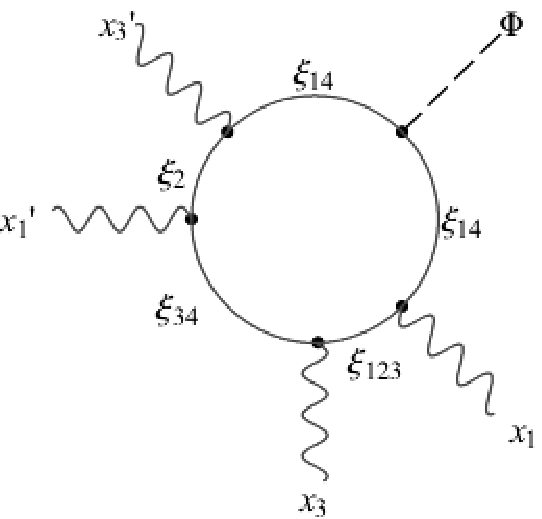}
\caption{ $\Phi (\Vec{i i})\to \psi \mC\psi,\phi$ type.} 
\label{scalarPhi14a}
\end{center}
\end{minipage}
\begin{minipage}[b]{0.47\linewidth}
\begin{center}
\includegraphics[width=4cm,angle=0,clip]{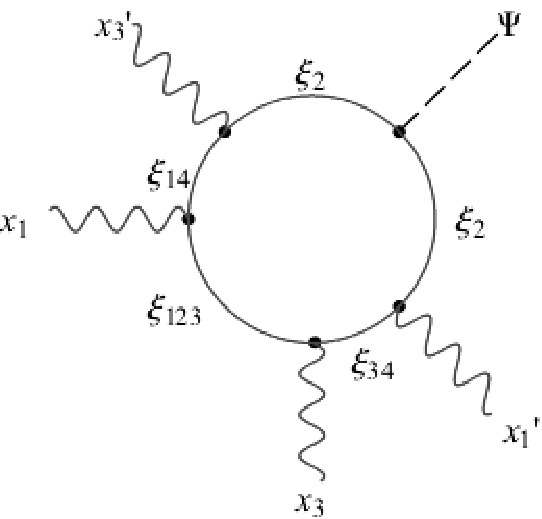}
\caption{ $\Psi (\Vec{j j})\to \phi \phi, \mC\psi$ type.} 
\label{scalarPsi123}
\end{center}
\end{minipage}
\hfill
\begin{minipage}[b]{0.47\linewidth}
\begin{center}
\includegraphics[width=4cm,angle=0,clip]{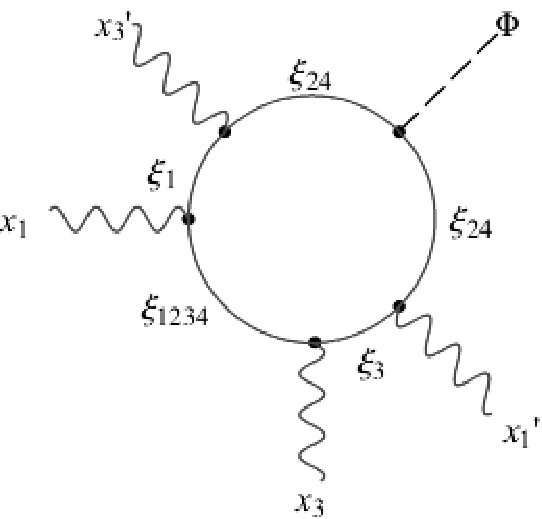}
\caption{ $\Phi (\Vec{j j})\to \psi \psi, \mC\phi$ type.} 
\label{scalarPhi1234}
\end{center}
\end{minipage}
\end{figure}
\begin{figure}[htb]
\begin{minipage}[b]{0.47\linewidth}
\begin{center}
\includegraphics[width=4cm,angle=0,clip]{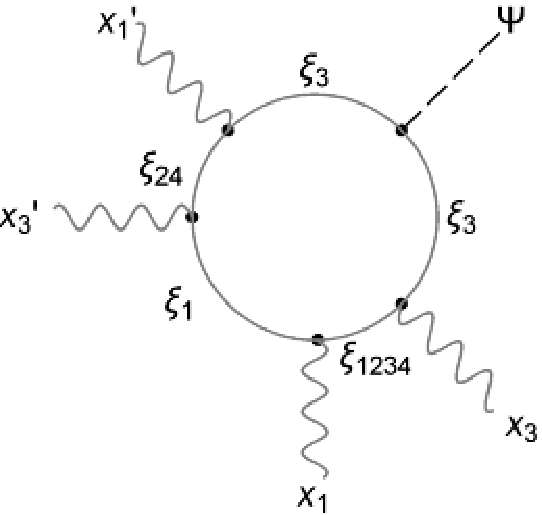}
\caption{ $\Psi (\Vec{k k})\to \phi \mC\phi,\psi$ type.} 
\label{scalarPsi234b}
\end{center}
\end{minipage}
\hfill
\begin{minipage}[b]{0.47\linewidth}
\begin{center}
\includegraphics[width=4cm,angle=0,clip]{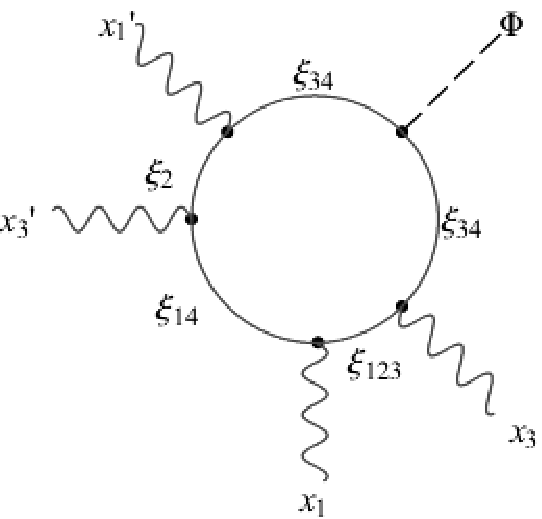}
\caption{ $\Phi (\Vec{k k})\to \psi \mC\psi,\phi$ type.} 
\label{scalarPhi14b}
\end{center}
\end{minipage}
\begin{minipage}[b]{0.47\linewidth}
\begin{center}
\includegraphics[width=4cm,angle=0,clip]{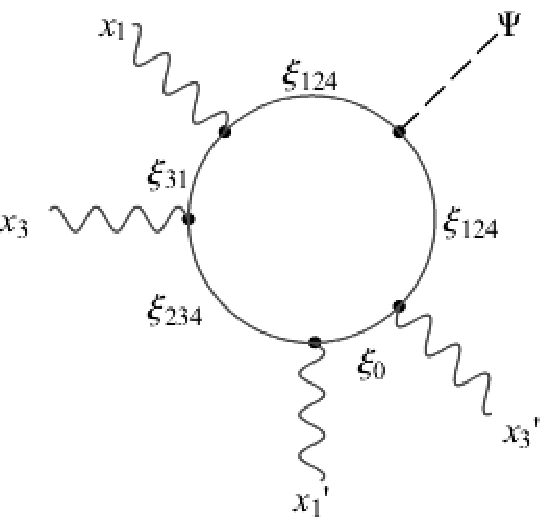}
\caption{ $\Psi(\Vec{k k})\to \mC\phi\phi, \mC\psi$ type.} 
\label{scalarPsi234a}
\end{center}
\end{minipage}
\hfill
\begin{minipage}[b]{0.47\linewidth}
\begin{center}
\includegraphics[width=4cm,angle=0,clip]{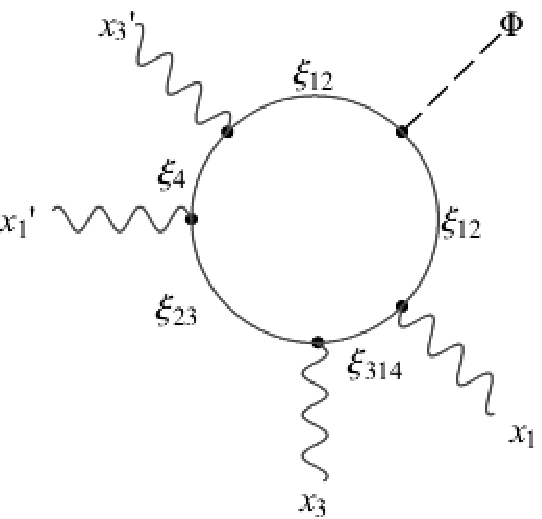}
\caption{ $\Phi (\Vec{k k})\to\psi \mC\psi,\phi$ type.} 
\label{scalarPhi23}
\end{center}
\end{minipage}
\end{figure}
\begin{figure}[htb]
\begin{minipage}[b]{0.47\linewidth}
\begin{center}
\includegraphics[width=4cm,angle=0,clip]{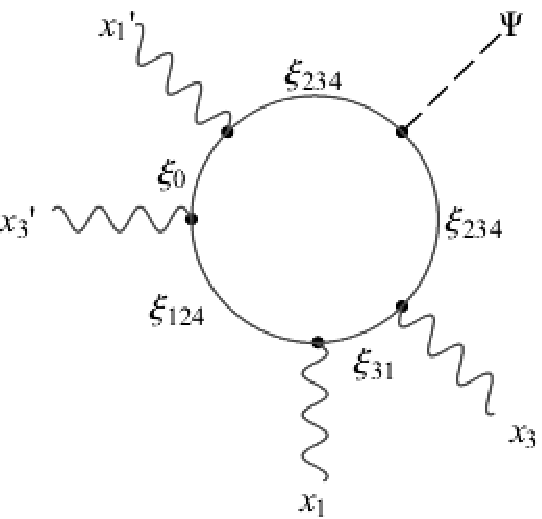}
\caption{ $\Psi(\Vec{i i})\to \phi \mC\phi, \mC\psi$ type.} 
\label{scalarPhi124}
\end{center}
\end{minipage}
\hfill
\begin{minipage}[b]{0.47\linewidth}
\begin{center}
\includegraphics[width=4cm,angle=0,clip]{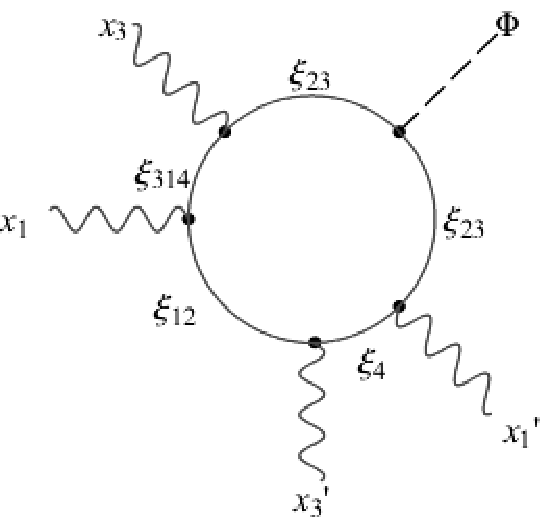}
\caption{ $\Phi (\Vec{i i })\to \mC\psi\psi, \phi$ type.} 
\label{scalarPsi12}
\end{center}
\end{minipage}
\begin{minipage}[b]{0.47\linewidth}
\begin{center}
\includegraphics[width=4cm,angle=0,clip]{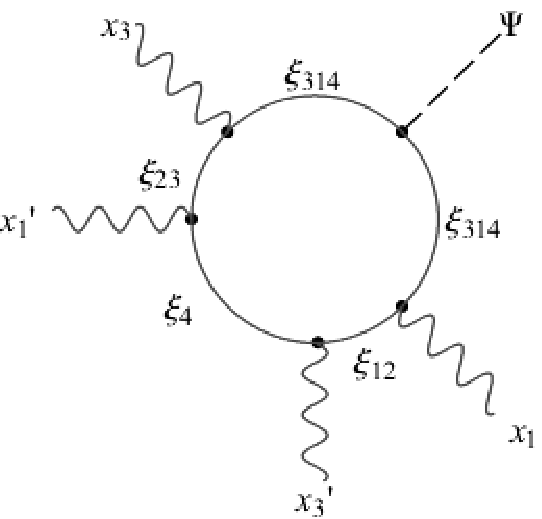}
\caption{ $\Psi(\Vec{j j})\to \mC\phi \mC\phi,\psi $ type.} 
\label{scalarPhi4}
\end{center}
\end{minipage}
\hfill
\begin{minipage}[b]{0.47\linewidth}
\begin{center}
\includegraphics[width=4cm,angle=0,clip]{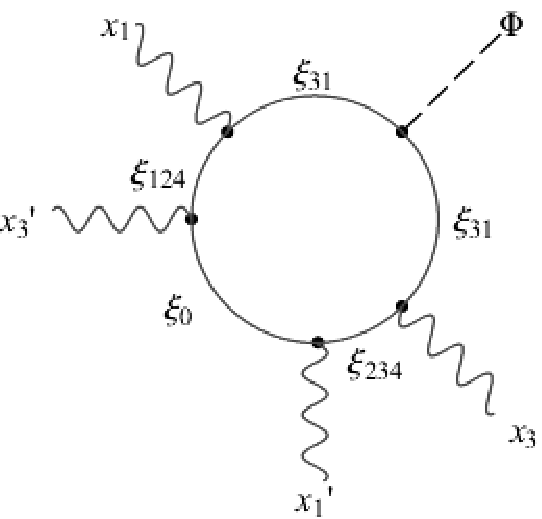}
\caption{ $\Phi (\Vec{j j})\to \mC\psi \mC\psi, \phi$ type.} 
\label{scalarPsi0}
\end{center}
\end{minipage}
\end{figure}

\subsection{Higgs boson decay into $\gamma\gamma$ via $\ell\bar\ell\ell\bar\ell$}
From the diagrams of scalar boson decay into four photons, we interchange the two emitted photons near the $\Psi$ or $\Phi$ bases to a lepton which posesses the same Dirac basis $\xi$ as the quark on the loop in the process of the decays. We interchange the other emitted photons to leptons, whose Dirac basis $\xi$ can be defined when the photons on the original lepton loops are specified.  We choose the specification of $\xi$ on the other size of $\Psi$ or $\Phi$, such that the two emitted leptons are on a line along $\Vec{i}, \Vec{j},$ or $\Vec{k}$ which corresponds to the lepton pair production, or on a line $I$ which corresponds to emission of the energy.

Since the two photons decay occur mainly from conversions of 2 lepton pairs\cite{ATLAS14a,ATLAS14b}, we study
\[
 H^0\to Z\bar Z\to \ell\bar\ell\ell\bar\ell\to\gamma\gamma .
\]
The scalar bosons $\Psi$ or $\Phi$ decays into two vector bosons $Z \bar Z$, and each $Z$ and $\bar Z$ emit two leptons. We assume that lepton pairs which are along a line interact by exchanging a photon $E$ with momenta $x_1$ or $x_3$ or $E'$ with momenta $x_1'$ or $x_3'$.  The directions of the two photons are arbitrary, if they are not parallel.  
When the interacting lepton pairs are $\phi \mC\phi, \psi \mC\psi$ or $\phi\psi, \mC\phi \mC\psi$, since $\phi$ and $\psi$ are relatively anti-particles, each pair convert to a $\gamma$.

In standard QCD, photons are different from gluons and information on triality is not expected, but when photons  are produced from pair annihilation, one could imagine arrival of information of triality on photons.

The distinction of $\Psi$ type scalar bosons and the $\Phi$ type scalar bosons should
be possible by observations of polarizations and distributions of produced photons or leptons.
From the diagrams of scalar boson decay into 4 photons,  we interchanged the first produced photon pairs and subsequently produced lepton pairs, and searched a lepton that interacts with the interchanged photon and interchanges themselves to the anti-lepton of the first produced lepton.

There are 8 diagrams of $\Psi$ decay into 2 lepton pairs and 8 diagrams of $\Phi$ decay into 2 lepton pairs, when directions of the momenta of photons exchanged between the two lepton pairs are chosen to be along the $x$ direction and along the $z$ direction. 
There are 4 diagrams in which $\xi_{1234}I$ and $\xi_{123}I$ energy pair and $\xi_2\sigma_y\sim\xi_2\Vec j$ and $\xi_{24}\sigma_y\sim \xi_{24}\Vec j$ lepton pairs are produced, and 4 diagrams in which $\xi_0 I$ and $\xi_4 I$ energy pair and $\xi_{314}\sigma_y\sim\xi_{314}\Vec j$ and $\xi_{31}\sigma_y\sim\xi_{31}\Vec j$ lepton pairs are produced. Other 8 diagrams are 2 lepton pairs production.

When a $\Psi$ decays into $\xi_{4}\mC{\xi_{4}}$ and produced leptons are $\xi_{23}\sigma_x\sim \xi_{23}\Vec i$ and $\xi_{12}\sigma_z\sim \xi_{12}\Vec k$, and the secondary leptons are $\xi_{234}\sigma_x\sim\xi_{234}\Vec i$ and $\xi_{124}\sigma_z\sim\xi_{124}\Vec k$ and the lepton that joins $\xi_{234}\sigma_x\sim\xi_{234}\Vec i$ and $\xi_{124}\sigma_z\sim\xi_{124}\Vec k$ is $\xi_0 I$, we specify the process as
\[
\Psi (11)\to \mC\phi \mC\psi, \mC\phi \mC\psi \quad ( or \quad \Psi(11)\to \phi\psi, \phi\psi)
\]
since $\xi_{23}\Vec i$ belongs to the $\mC\phi$ group and $\xi_{234}\Vec i$ belongs to the $\mC\psi$ group, and similarly $\xi_{12}\Vec k$ belongs to the $\mC\phi$ group. 

\pagebreak[4]
\begin{figure}[htb]
\begin{minipage}[b]{0.47\linewidth}
\begin{center}
\includegraphics[width=4cm,angle=0,clip]{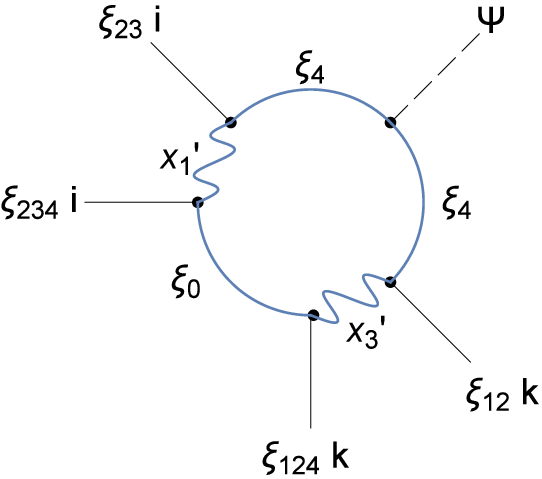}
\caption{Scalar boson decay into 4 leptons.  $\Psi (11)\to \mC\phi \mC\psi, \mC\phi \mC\psi$ type. }
\label{lscalarPsi244}
\end{center}
\end{minipage}
\hfill
\begin{minipage}[b]{0.47\linewidth}
\begin{center}
\includegraphics[width=4cm,angle=0,clip]{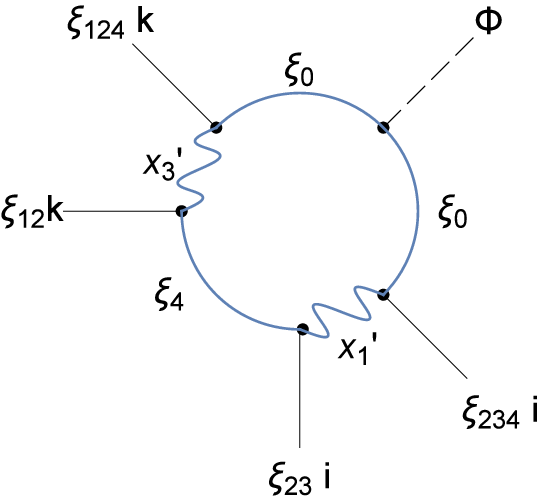}
\caption{Scalar boson decay into 4 leptons. $\Phi (11)\to \mC\psi \mC\phi,\mC\psi \mC\phi$ type. }
\label{lscalarPhi240 }
\end{center}
\end{minipage}
\begin{minipage}[b]{0.47\linewidth}
\begin{center}
\includegraphics[width=4cm,angle=0,clip]{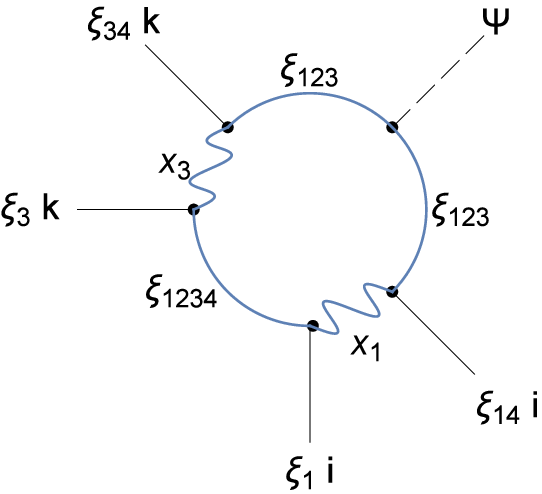}
\caption{ $\Psi (11)\to \phi\psi, \phi\psi$ type.} 
\label{lscalarPsi314}
\end{center}
\end{minipage}
\hfill
\begin{minipage}[b]{0.47\linewidth}
\begin{center}
\includegraphics[width=4cm,angle=0,clip]{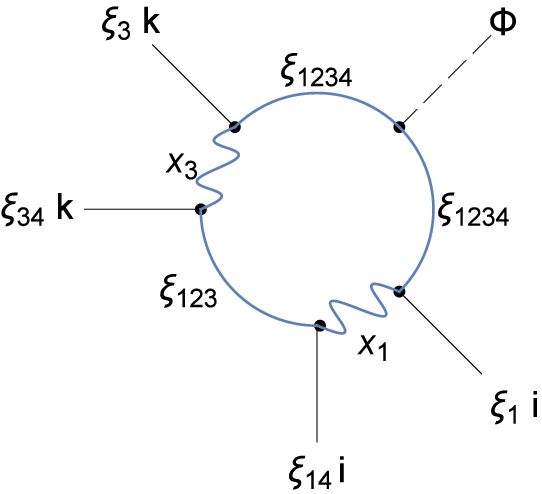}
\caption{ $\Phi (11)\to \psi\phi ,\psi\phi$ type.} 
\label{lscalarPhi4a}
\end{center}
\end{minipage}
\begin{minipage}[b]{0.47\linewidth}
\begin{center}
\includegraphics[width=4cm,angle=0,clip]{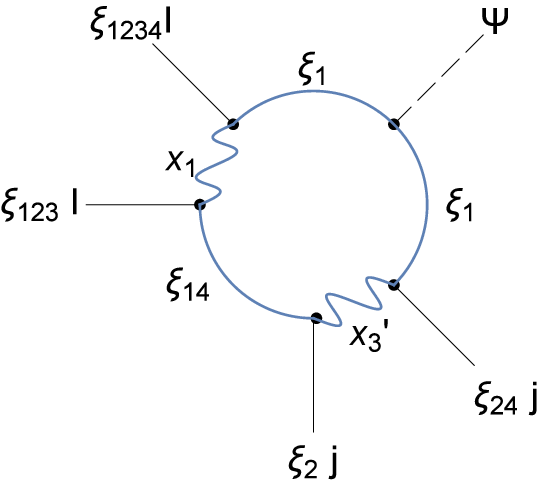}
\caption{ $\Psi (\Vec{i i})\to \mC\phi \mC\psi,\mC\phi \mC\psi$ type.} 
\label{lscalarPsi3}
\end{center}
\end{minipage}
\hfill
\begin{minipage}[b]{0.47\linewidth}
\begin{center}
\includegraphics[width=4cm,angle=0,clip]{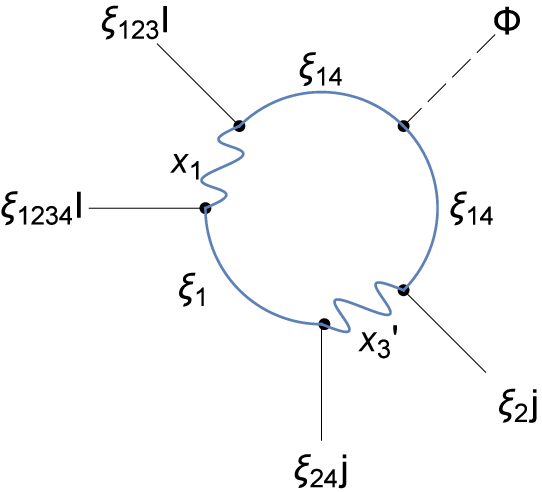}
\caption{ $\Phi (\Vec{k k})\to \mC\psi \mC\phi,\mC\psi \mC\phi$ type.} 
\label{lscalarPhi14a}
\end{center}
\end{minipage}
\begin{minipage}[b]{0.47\linewidth}
\begin{center}
\includegraphics[width=4cm,angle=0,clip]{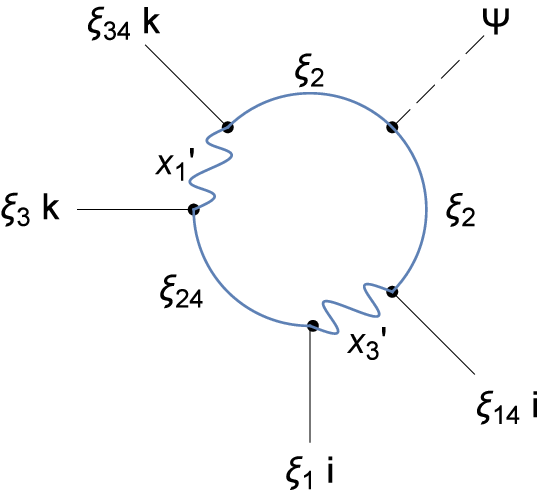}
\caption{ $\Psi (\Vec{j j})\to \psi\phi,\psi\phi$ type.} 
\label{lscalarPsi123}
\end{center}
\end{minipage}
\hfill
\begin{minipage}[b]{0.47\linewidth}
\begin{center}
\includegraphics[width=4cm,angle=0,clip]{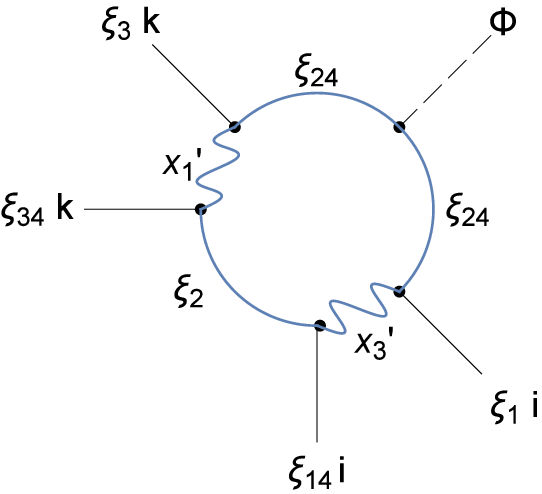}
\caption{ $\Phi (\Vec{j j})\to \psi\phi,\psi\phi$ type.} 
\label{lscalarPhi1234}
\end{center}
\end{minipage}
\end{figure}
\begin{figure}
\begin{minipage}[b]{0.47\linewidth}
\begin{center}
\includegraphics[width=4cm,angle=0,clip]{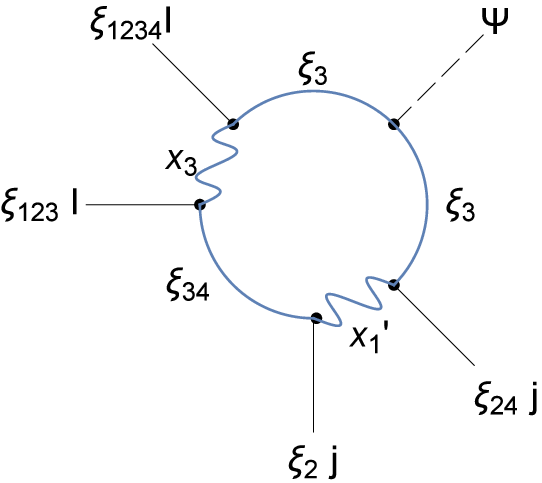}
\caption{ $\Psi (\Vec{k k})\to \mC\phi \mC\psi, \psi\phi$ type.} 
\label{lscalarPsi234b}
\end{center}
\end{minipage}
\hfill
\begin{minipage}[b]{0.47\linewidth}
\begin{center}
\includegraphics[width=4cm,angle=0,clip]{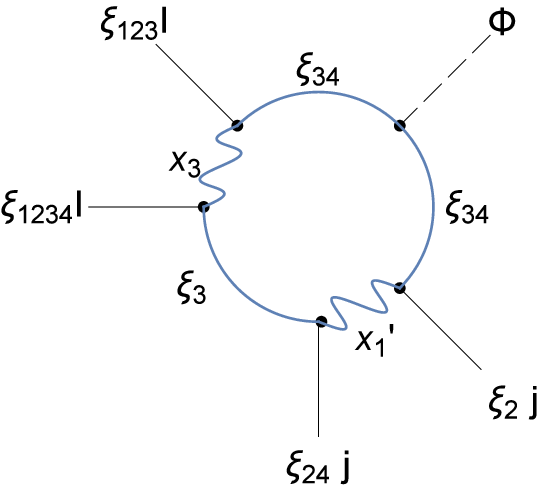}
\caption{ $\Phi (\Vec{k k})\to \mC\psi \mC\phi, \phi\psi$ type.} 
\label{lscalarPhi14b}
\end{center}
\end{minipage}
\begin{minipage}[b]{0.47\linewidth}
\begin{center}
\includegraphics[width=4cm,angle=0,clip]{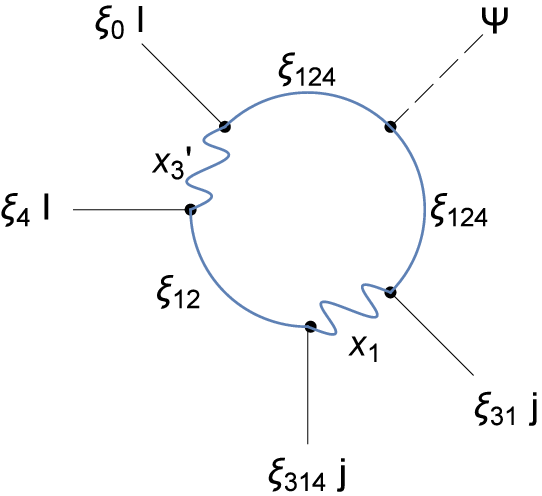}%
\caption{ $\Psi(\Vec{k k})\to \phi\psi, \mC\psi \mC\phi$ type.} 
\label{lscalarPsi234a}
\end{center}
\end{minipage}
\hfill
\begin{minipage}[b]{0.47\linewidth}
\begin{center}
\includegraphics[width=4cm,angle=0,clip]{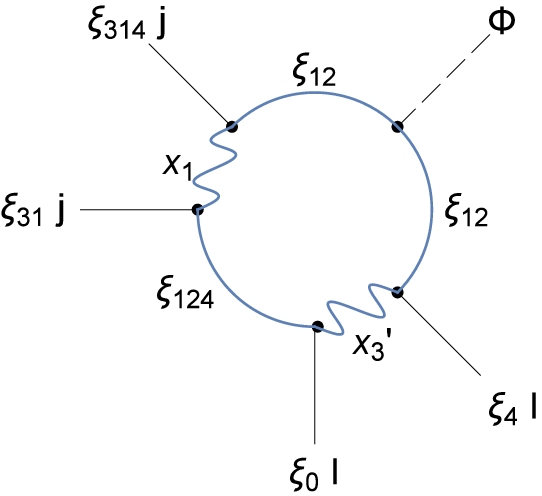}
\caption{ $\Phi (\Vec{k k})\to \mC\psi \mC\phi,\mC\psi \mC\phi$ type.} 
\label{lscalarPhi23}
\end{center}
\end{minipage}
\begin{minipage}[b]{0.47\linewidth}
\begin{center}
\includegraphics[width=4cm,angle=0,clip]{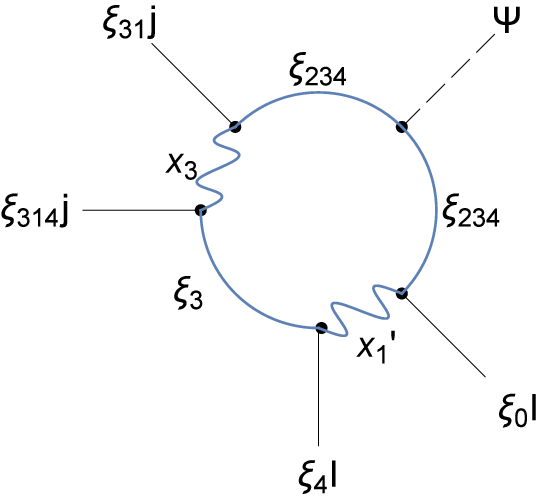}
\caption{ $\Psi(\Vec{i i})\to \mC\phi \mC\psi, \psi\phi$ type.} 
\label{lscalarPhi124}
\end{center}
\end{minipage}
\hfill
\begin{minipage}[b]{0.47\linewidth}
\begin{center}
\includegraphics[width=4cm,angle=0,clip]{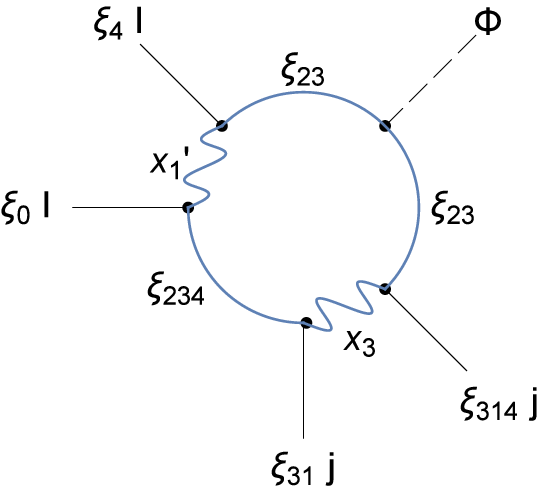}
\caption{ $\Phi (\Vec{i i })\to \psi\phi,\mC\phi \mC\psi$ type.} 
\label{lscalarPsi12}
\end{center}
\end{minipage}
\begin{minipage}[b]{0.47\linewidth}
\begin{center}
\includegraphics[width=4cm,angle=0,clip]{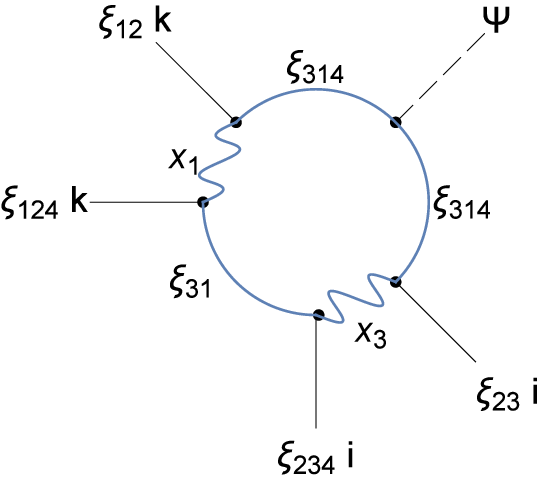}
\caption{ $\Psi(\Vec{j j})\to \mC\phi \mC\psi,\mC\phi \mC\psi$ type.} 
\label{lscalarPhi4}
\end{center}
\end{minipage}
\hfill
\begin{minipage}[b]{0.47\linewidth}
\begin{center}
\includegraphics[width=4cm,angle=0,clip]{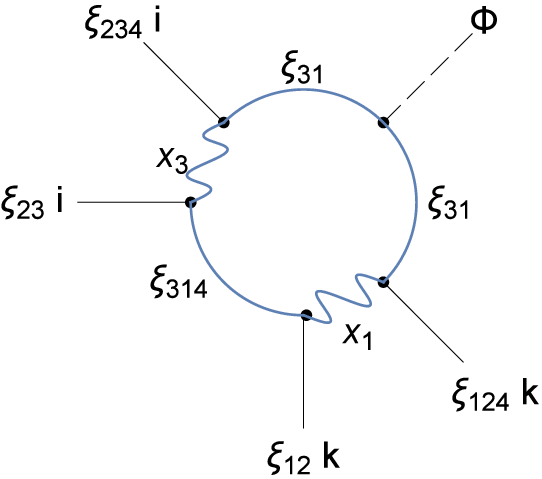}
\caption{ $\Phi (\Vec{j j})\to \psi\phi,\psi\phi$ type.} 
\label{lscalarPsi0}
\end{center}
\end{minipage}
\end{figure}

We studied Higgs boson decay into photons assuming production of lepton pairs expressed by Cartan's spinors. After the emission of leptons along the $x$ axis and the $z$ axis or along the $y$ axis and energy emission, we assumed two photons are emitted along the $x$ axis and the $z$ axis, and we found 8 diagrams from $\Psi$ and 8 diagrams from $\Phi$. Emission of photons along the $x$ axis and $y$ axis and along the $y$ axis and the $z$ axis are the same. The lepton pairs can decay into a photon, and the process can be detected as decay into $\gamma\gamma$. 

\pagebreak[4]
\section{Dark Matter and the triality symmetry}
Incorporation of Dirac fields $\phi(A,B)$ and $\psi(C,D)$ introduces 8 dimensional vector field $(x^1,\cdots,x^4,$ ${x^1}',\cdots,{x^4}')$ which has the octonion symmetry
\[
\Vec O=\Vec R\oplus \Vec R^{0,7}.
\]
The automorphism group of $\Vec O$ is a subgroup of the exceptional Lie group $G_2$ which has the triality symmetry.
 
The Clifford group ${\Vec \Gamma}_{p,q}$ is defined as\cite{Lo93,Lo01} 
\[
{\Vec \Gamma}_{p,q}=\{s\in Cl_{p,q}^+\cup Cl_{p,q}^-|\forall{\Vec x}\in {\Vec R}^{p,q}, s{\Vec x}s^{-1}\in {\Vec R}^{p,q}\}. 
\]
A normalized subgroup of Clifford group is defined as $\Vec{Pin}$:
\[
\Vec{Pin}(p,q)=\{ s\in\Gamma_{p,q}| s\tilde s=\pm 1\},
\]
where $\tilde s$ is a reversion of $s$, and when $s\tilde s$ is restricted to +1, $\Vec\Gamma_{p,q}$ is called 
$\Vec{Spin}$.

In the Clifford Algebra $C\ell_{0,7}$ of ${\Vec R}^{0,7}$, one can define a trivector
\[
{\Vec v}= {\Vec e}_{124}+{\Vec e}_{235}+{\Vec e}_{346}+{\Vec e}_{457}+{\Vec e}_{561}+{\Vec e}_{672}+{\Vec e}_{713}
\]
and $\Vec w={\Vec v}{\Vec e}_{12\cdots 7}^{-1}\in \wedge^4{\Vec R}^{0,7}$.

The triality can be viewed as a property of the spin group
\[
\$ pin(8)=\{u\in C\ell_{0,7}|u\bar u=1,\quad {\rm for\quad all }\quad x\in \$\Vec R^8\quad {\rm also}\quad ux\hat u^{-1}\in \$\Vec R^8\}
\]
and the minimal left ideal $C\ell_{1,7}\frac{1}{8}(1+{\Vec w})\frac{1}{2}(1\pm{\Vec e}_{12\cdots 7})$ and primitive idempotents
\[
f=\frac{1}{8}(1+{\Vec w})\frac{1}{2}(1-{\Vec e}_{12\cdots 7})\quad {\rm and}\quad 
\hat f=\frac{1}{8}(1+{\Vec w})\frac{1}{2}(1+{\Vec e}_{12\cdots 7}).
\]
Triality transformation corresponds to the multiplication of $1 ,f$ and $\hat f$.


Clifford Algebra $Cl_3^+$ which has bases $\Vec R\oplus\Vec R^3\oplus \wedge^2\Vec R^3\oplus\wedge^3\Vec R^3$ can make a $\Vec{Spin}(3)$ which is expressed by quaternions: $\Vec H=\{1,\Vec i,\Vec j,\Vec k\}$, and an extension with a new imaginary unit $\Vec l$, $\Vec H\oplus \Vec H\Vec l $ is expressed by an octonion $\Vec O=\Vec R^8$ instead of $\Vec R\oplus \Vec R^7$.

All automorphisms of $\Vec{Spin}(n), n\ne 8$ are of the form $u\to sus^{-1}$ where $s\in \Vec{Pin}(n)$, but the
group $\Vec{Spin}(8)$ has exceptional automorphisms, which permute the non-identity elements -1,${\bf e}_{12\cdots 8}, -{\Vec e}_{12\cdots 8}$ in the center of $\Vec{Spin}$:
\[
-1\to{\Vec e}_{12\cdots 8}\to -{\Vec e}_{12\cdots 8}\to -1.
\]
This automorphism of order three is called triality automorphism\cite{Lo93,Lo01,He86}. 

The $\Vec{Spin}(8)$ is constructed  from $Cl_8^+\in \Vec R^8$, ${Cl_8^+}\Vec f_+$ and ${Cl_8^+}\Vec f_-$ where $\Vec f_\pm=\frac{1}{8}(1+\Vec w)\frac{1}{2}(1\pm \Vec e_{12\cdots 8})$, and it has the triality symmetry.  
We want to assign the elementary particles and the field, the symmetry of $\Vec{Spin}(8)$.

In the Cartan's spinor theory based on Clifford algebra\cite{Cartan66}, leptons have two superpartners which can be hidden by our electromagnetic probes.  Leptons $(A,B,C,D$) and vector field $(E, E')$ are transformed by $G_{23},G_{12},G_{13}$ and $G_{123},G_{132}$. In the standard model, leptons consist of $e,\mu,\tau$, and quarks consist of $u,d,s,c,b,t$ flavors and each component has $r,b,g$ colors. Electromagnetic probes have triality selection rules, and detect $E$ or $E'$ in the same triality sector as that of electrons in the detector have.


When electromagnetic probes do not detect electromagnetic waves 
$(E, E')$ transformed by $G_{12},G_{13},$ $G_{123}$ and $G_{132}$, then 4/6 of the electromagnetic waves in the universe will appear as dark energy.

The transformation ${G_{23}}$ interchanges two components of the four component spinor, or it transforms  matter to anti-matter.  Then  5/6 of the matter in the universe appear different from matter in our universe and appear as  dark matter.  The Wilkinson Microwave Anisotropy Probe (WMAP ) space craft confirms that almost five times more dark matter (24\%) than the normal matter (4.6\%) are observed \cite{Xe12,Blitz11,NASA14}. 

 We considered three neutrinos in different triality sectors interacting with each other and produced one heavy and two degenerate light neutrinos, which are $\nu_e, \nu_\mu$ and $\nu_\tau$\cite{SF13b}.  Their lepton partners,  $ e, \mu$ and $\tau$ are sensitive to the flavors and the triality of electromagnetic waves, but blind to the triality of quarks and gluons.  If electromagnetic waves from different triality sectors cannot be detected by electromagnetic probes in our world, we can understand the presence of dark matter. 

 Assuming the triality selection rules of octonions, dark matter is interpreted as matter emitting photons in a different triality sector than that of electromagnetic probes in our world.
Satellite galaxies of Milky Way are a promising target for dark matter searches in gamma rays. The Milky Way is expected to produce gamma rays in different triality.
 The dark matter is an object whose light cannot be detected on our electromagnetic detectors, but it should be detected through gravitational influences.

\vskip 0.5 true cm
{\small
The author thanks Professor Bitbol for sending me his papers and valuable information, and Dr. Serge Dos Santos for discussions.}

\begin{thebibliography}{99}
%
%
\bibitem{Ge07}
Georgi H.(2007), Unparticle Physics, \PRL{\bf 98}, 221601.
\bibitem{Higgs64}
 Higgs P.W. (1964), Broken symmetries and the mass of gauge bosons, \PRL{\bf 13}, 508.
\bibitem{EWPD12} Espinosa J.R., Grojean C., M\"uhlleitner M., and Trott M. (2012), Fingerprinting Higgs Suspects at the LHC,  arXiv:1202.3697v2[hep-ph]
\bibitem{Cartan66}
Cartan \'E. (1966), {\it The theory of Spinors}, Dover Pub., p.118.
\bibitem{SF12a}
 Furui S.(2012a), Fermion Flavors in Quaternion Basis and Infrared QCD, Few Body Syst. {\bf 52}, 171-187(2012); 
\bibitem{SF12b}
 Furui S.(2012b), The Magnetic Mass of Transverse Gluon, the B-Meson Weak Decay Vertex and the Triality Symmetry of Octonion, Few Body Syst. {\bf 53}, 343.
\bibitem{SF13a} 
Furui S. (2013), Axial anomaly and triality symmetry of octonion, Few Body Syst. DOI 10.1007/s0061-013-0719-9, arXiv:1301.2095 [hep-ph].
\bibitem{SF13b} 
Furui S. (2014), Axial anomaly and triality symmetry of leptons and hadrons, Few Body Syst. {\bf 55}, 1083, arXiv:1304.3776 [hep-ph].
\bibitem{Xe12} 
Aprile E. et al (XENON100 Collaboration) (2012), The XENON100 Dark Matter Experiment, Astropart. Phys. 35, 573, arXiv:1107.2155[astro-ph.IM].
\bibitem{PS95}
Peskin M.E. and Schroeder D. (1995), An Introduction to Quantum Field Theory, Perseus Books.
\bibitem{BBJ81} 
Becher P., B\"ohm M. and Joos H. (1981), Eichtheorien, Teubner Studienb\"ucher, Teubner.
\bibitem{SF15}
Furui S. (2015), Cartan's Supersymmetry and the violation of CP symmetry, arXiv:1505.05830.
\bibitem{BaBar09}
Aubert B. et al. (BaBar Collaboration) (2009), Measurement of Time-Dependent CP Asymmetry in $B^0\to c\bar cK^{(*)0}$ Decays, Phys. Rev. D{\bf 79}, 072009.
\bibitem{Bitbol85}
Bitbol M. (1986), Time Symmetry and Quantum Measurements, Phys. Lett. A {\bf 115} 357.
\bibitem{Adler65} 
Adler S.L. (1965), Consistency Conditions on the Strong Interactions Implied by a Partially Conserved Axial-Vector Current II, Phys. Rev.{\bf 139}, B1638-1643.
\bibitem{AB69} 
 Adler S.L. and Bardeen W.A. (1969),  Absence of higher-order corrections in the anomalous axial-vector divergence equation, Phys.Rev.{\bf 182},1517.
\bibitem{GL84} 
Gasser J. and Leutwyler H. (1984), Chiral Perturbation Theory to One Loop, Ann.Phys. {\bf 158},142-210.
\bibitem{CMS11} The CMS collaboration (2011), CMS Physics Analysis Summary, CMS PAS HIG-11-011.
\bibitem{ATLAS12} 
The ATLAS collaboration (2012), ATLAS NOTE, ATLAS-CONF-2012-09. 
\bibitem{ATLAS11} 
The ATLAS collaboration (2011),  Search for the Higgs boson in the $H\to WW^{(*)}\to l^+\nu l^-\bar\nu$ decay charannel in $pp$ collisions at $\sqrt s=7$ TeV with the ATLAS detector, \PRL{\bf 108}.111802, arXiv:1112.2577. 
\bibitem{CMS13}
 The CMS collaboration (2013),  Search for the Standard Model Higgs boson in the H to WW to $l\nu jj$decay dhannel in $pp$ collisions at the LHC, CMS Collaboration, CMS PAS HIG-13-027.
\bibitem{ATLAS14a} 
Aad G. et al. (ATLAS Collaboration)(2014), Measurement of the Higgs boson mass from the $H\to\gamma\gamma$ and $H\to ZZ^*\to 4l$ channels in $PP$ collisions at center-of-mass energies of 7 and 8 TeV with the ATLAS detector, \PRD{\bf 90},052004.
\bibitem{ATLAS14b} The ATLAS collaboration (2014), Electron and photon energy calibration with the ATLAS detector using LHC Run 1 data, arXiv:1407.5063v2.
\bibitem{DLM00} 
Dobrescu B.A. , Landsberg G.  and Maltchev K.T. (2001), Higgs Boson Decays to CP-odd Scalars at the Tevatron and Beyond, \PRD {\bf 63}, 075003, arXiv.hep-ph/0005308v1.
\bibitem{CFW07} 
Chang S. , Fox P.J., and Weiner N. (2007), Visible Cascade Higgs Decays to Four Photons at Hadron Colliders, \PRL{\bf 98},111802. 
\bibitem{DHL85} 
Donoghue J.F. , Holstein B.R. , and Lin Y.-C.R. (1985), Chiral Loops in $\pi^0,\eta^0\to\gamma\gamma$ and $\eta-\eta'$ Mixing, \PRL{\bf 55},2766-2769.
\bibitem{BBC88} 
Bijens J. , Bramon A.  and Cornet F. (1988), Pseudoscalar Decays into Two Photons in Chiral Perturbation Theory, \PRL{\bf 61}, 1453.
\bibitem{KaLe00} 
Kaiser R.  and Leutwyler H. (20001), Large $N_c$ in chiral perturbation theory, Eur. Phys. J. {\bf C17},623, arXiv:hep-ph/0007101
\bibitem{BoWe01} 
Borasoy B. and Wetzel S. (2001), $U(3)$ chiral perturbation theory with infrared regularization, \PRD{\bf 63}, 074019.
\bibitem{BeBo01} 
Beisert N. and Borasoy B. (2001), $\eta-\eta'$ mixing in $U(3)$ chiral perturbation theory, Eur. Phys. J. {\bf A11},329, arXiv:hep-ph/0107175
\bibitem{BoNi03} 
Borasoy B. and Nissler R. (2004), Two-photon decay of $\pi^0,\eta$ and $\eta'$, Eur. Phys. J. {\bf A19}, 367, arXiv:hep-ph/0309011
\bibitem{Shore01} 
Shore G.M. (2002),  $\eta'(\eta)\to\gamma\gamma$: A Tale of Two Anomalies,  Phys. Scripta {\bf T99} , 84, arXiv:hep-ph/0111165.
\bibitem{MOU13} Michael C. , Ottnad K. , and Urbach C. (2013),  $\eta$ and $\eta'$ mixing from Lattice QCD, \PRL {\bf 111}, 18, arXiv:1310.1207[hep-lat] 
\bibitem{SHKM10} 
Sibirtsev A. , Haidenbauer J. , Krewald S.  and Meissner U.-G. (2010), Analysis of recent eta photoproduction data, Eur. Phys. J. {\bf A46} 359-371, arXiv:1007.3140v2[nucl-th].
\bibitem{Hab00} 
Haberzettl H. (2000), Pion photo-and electroproduction and partially-conserved axial current, \PRL{\bf 85},3576, arXiv:nucl-th/0005016v3.
\bibitem{CBTA09} The CBELSA/TAPS Collaboration (2009), Photoproduction of $\eta$ and $\eta'$ Mesons off Protons, \PRC {\bf 80}, 055202, arXiv:0909.1248v2[nucl-ex].
\bibitem{Lo93} 
Lounesto P. (1993), Clifford algebras and Hestenes spinors, Foundation of Physics, {\bf 23}, 1203-1237.
\bibitem{Lo01} 
Lounesto P. (2001), in {\it Clifford Algebras and Spinors} 2nd ed. ,Cambridge University Press.
\bibitem{He86} 
Hestenes D. (1986), Clifford Algebras and their Applications in Mathematical Physics, Reidel, Dordrecht/Boston , 321
\bibitem{Blitz11} Blitz L.(2011) , Dark Matter, Scientific American, p.38 October.
\bibitem{NASA14} NASA Science Astrophysics (2014), Dark Energy, Dark Matter.
\end{thebibliography}


\end{document}